\documentclass[letterpaper,twocolumn,10pt]{article}
\usepackage{usenix-2020-09}

\usepackage{tikz}
\usepackage{amsmath}
\usepackage{enumitem}
\usepackage[normalem]{ulem}
\usepackage{graphicx}
\usepackage{subfigure}
\usepackage{amsfonts}
\usepackage{booktabs}   
\usepackage{siunitx}
\usepackage{appendix}
\usepackage{pifont}
\usepackage{multirow}   
\usepackage{makecell}   
\usepackage[ruled,vlined,linesnumbered]{algorithm2e}

\begin{document}

\pagestyle{empty}
\thispagestyle{empty}
\date{}

\title{\Large \bf RIGEL: Real-time Optical Anomaly Diagnosis with Stateful In-Network Inference based on Distributed On-switch GNNs}

\author{Zhen Wei, Yidong Wang, Yufan Zhu, Xuefeng Yan, Binjun Tang, Xiaoliang Chen, and Zuqing Zhu\\
School of Information Science and Technology, University of Science and Technology of China, Hefei, China\\
Email: {zqzhu@ieee.org}
}

\maketitle

\begin{abstract}

The recent booming of data-intensive applications has complicated optical network management, making real-time optical anomaly diagnosis a must-have feature. However, existing approaches are mostly based on centralized data analytics and thus can hardly avoid the latency and overhead due to message exchanges between data and control planes. In this work, we propose and prototype RIGEL, which, to the best of our knowledge, is the first real-time optical anomaly diagnosis system that realizes stateful distributed in-network inference through collaborative graph neural networks (GNNs) on Tofino switches. The system is designed to be fully in-network, and a software-hardware co-design is proposed to preprocess high-dimensional spectral data for being suitable for hardware-based in-network inference. Next, we first develop an effective model to combine an autoencoder with a GraphSAGE-based GNN, and then propose a generalizable method to adapt the model to Tofino switch. The effectiveness of RIGEL is showcased in a realistic packet-over-optical network testbed, verifying that it achieves highly accurate diagnosis to detect and locate optical anomalies timely and highlighting its benefits over the state-of-the-art methods.

\end{abstract}

\section{Introduction}

Nowadays, the rapid development of data-intensive applications, especially the rise of artificial intelligence (AI) and large language models (LLMs), has reshaped optical networks into dynamic and high-throughput underlay infrastructures with enhanced adaptability~\cite{Liu2023lightwave, Amir2024shale, Miao2023flexwan, Li2024opticgai}. 

Yet, this deep coupling with applications complicates the design of optical networks and pushes their operations to a fragile edge: minor and ephemeral anomalies can be amplified through cascaded processing, leading to unexpected service degradations~\cite{Miao2022OpTel, zhang2024expertise}, while in a wavelength-division multiplexing (WDM) based optical network, a fiber link normally carries tens of Tbps or more traffic, and thus a short downtime of even 1 ms brings more than 1 GB data loss. Hence, precise and real-time anomaly diagnosis becomes a critical feature for today's optical networks~\cite{Mellette2024RotorNet, zar2025toward}.

This diagnostic imperative, however, is not just about detecting the well-understood hard failures that will interrupt lightpaths immediately. The critical challenge lies in the proactive diagnosis of soft failures (\textit{i.e.}, the minor and subtle anomalies that might not tear down lightpaths right away but could lead to severe consequences over time), because they usually have complex and multi-variate root-causes that can hardly be detected by traditional methods~\cite{musumeci2019tutorial, sun2022digital, vela2017ber, lechowicz2025optimizing}. Therefore, researchers have switched to more sophisticated diagnosis techniques that adopt machine learning (ML)~\cite{silva2023confidentiality, chen2022automating, chen2022cooperative,lun2023gan}. 

Despite their effectiveness, these techniques usually need to fuse telemetry data across multiple optical nodes, which can be realized in either a centralized or distributed manner. The centralized architecture (\textit{e.g.}, software-defined networking (SDN)) is prone to single-point failures (\textit{i.e.}, the anomaly diagnosis cannot operate without the centralized controller) and bears additional latency and overheads for data analytics (\textit{i.e.}, a WDM-based optical network usually has a diameter of hundreds or even thousands of kilometers, meaning that the round-trip time between the controller and an optical node can be in milliseconds, and as the controller needs to handle other network control and management tasks as well, streaming telemetry data to it consistently can saturate its control channel and computing capability).

Recent approaches tried to address these issues by introducing distributed data analytics and offloading diagnosis tasks to servers placed close to optical nodes~\cite{sales2024disaggregated, lin2024scaling, musumeci2025failure}. Nevertheless, the servers bring in additional capital expenses and operational complexity, and they still can hardly avoid the delay and overhead intrinsic in hardware/software context switching, deviating from the real-time and in-network diagnosis required to prevent anomalies from deteriorating.

The limits of state-of-the-art methods call for a paradigm shift that can push data analytics for anomaly diagnosis closer to telemetry data sources and minimize remote computing. Although not originally designed for optical anomaly diagnosis, intelligent data plane (IDP) offers a promising platform for this vision by enabling line-rate ML inference directly on programmable hardware (SmartNICs~\cite{sanvito2018can, siracusano2020running}, FPGAs~\cite{zheng2022iisy, swamy2022taurus}, and P4-enabled switches based on Tofino ASICs~\cite{sapio2021scaling, zheng2021planter}). While the benefits of IDP have been predominantly studied in the packet domain (\textit{e.g.}, traffic classification~\cite{xavier2021programmable, siracusano2018deep} and fault detection~\cite{barradas2021flowlens, liu2021jaqen}), the line-rate processing with low latency will be more beneficial for real-time optical anomaly diagnosis, but it is still under-explored.

The implementation of ML inference in IDP began with the models naturally aligned with the capability of IDP. For example, those based on decision trees (DTs) were an early success for being directly realized with a pipeline of match-action tables (MATs)~\cite{xiong2019switches, busse2019pforest, zang2022p4pir, xie2023empowering, zhou2023efficient, butun2025dune}. The transition to more sophisticated neural networks (NNs), however, encountered a major hurdle due to the mismatch between their complex arithmetic operations and the simple primitives of IDP. This challenge was first addressed by simplifying ML models, leading to binarized NNs (BNNs)~\cite{siracusano2022re, xie2022mousika}. Then, recent breakthroughs considered more complex models, successfully offloading multi-bit convolutional NNs (CNNs) and recurrent NNs (RNNs) onto IDP~\cite{yan2024brain, zhang2025quark, zhang2025pegasus}. 

However, these prior studies mainly focused on processing packet headers in isolated switches. Therefore, they can hardly provide network-awareness through multi-switch collaboration, which is critical for optical anomaly diagnosis \cite{musumeci2025failure,chen2022cooperative}.

In this work, we bridge this critical gap by proposing and prototyping \textbf{RIGEL} (\uline{R}eal-time optical anomaly d\uline{I}agnosis with distributed on-switch \uline{G}raph n\uline{E}ura\uline{L} networks), which, to the best of our knowledge, is the first optical anomaly diagnosis system that achieves stateful distributed in-network inference with collaboration of graph neural networks (GNNs) on Tofino switches. 

RIGEL reuses the packet switch deployed aside each optical node (typical configuration of the packet-over-optical architecture) for cost saving, \textit{i.e.}, the cost of a Tofino switch is similar to that of a legacy one with the same throughput~\cite{Tofino}.

Although GNN is suitable for analyzing graph-structured telemetry data for optical anomaly diagnosis, offloading a GNN onto IDP is more challenging than other ML models due to its iterative logic, stateful neighbor aggregation, and the need to manage and aggregate high-dimensional feature embeddings within the data plane.

RIGEL resolves these challenges with breakthroughs in: 1) a GNN-based real-time optical anomaly diagnosis framework that is distributed and designed to be fully in-network (\S\ref{sec:Design Overview}), 2) a software-hardware co-design that reuses the computing capability of optical performance monitors (OPMs) to preprocess high-dimensional raw telemetry data through compressing and quantizing for being suitable for in-network inference in IDP (\S\ref{sec:preprocessing}), 3) a co-designed IDP-friendly model that combines an autoencoder (AE) for feature compression and a GraphSAGE for distributed in-network inference (\S\ref{sec:Two-stage}), and 4) a generalizable method that offloads IDP-incompatible computations to IDP pipelines by leveraging the neighborhood scoping of GraphSAGE~\cite{hamilton2017inductive} (\S\ref{sec:Multi-stage}).

The effectiveness of RIGEL is demonstrated through extensive experiments in a realistic network testbed built with optical components (wavelength-selective switches (WSS'), erbium-doped fiber amplifiers (EDFAs), and bandwidth-variable transponders (BVTs)) and IDP switches based on Tofino ASICs. The results show that RIGEL achieves highly accurate optical anomaly diagnosis, reducing data exchange overhead between data and control planes by more than three orders of magnitude and significantly accelerating diagnosis.

This work does not raise any ethical issues.

\begin{figure}[t] 
	\centering 
	\subfigure[In-band noise jamming]{
		\includegraphics[width=0.32\textwidth]{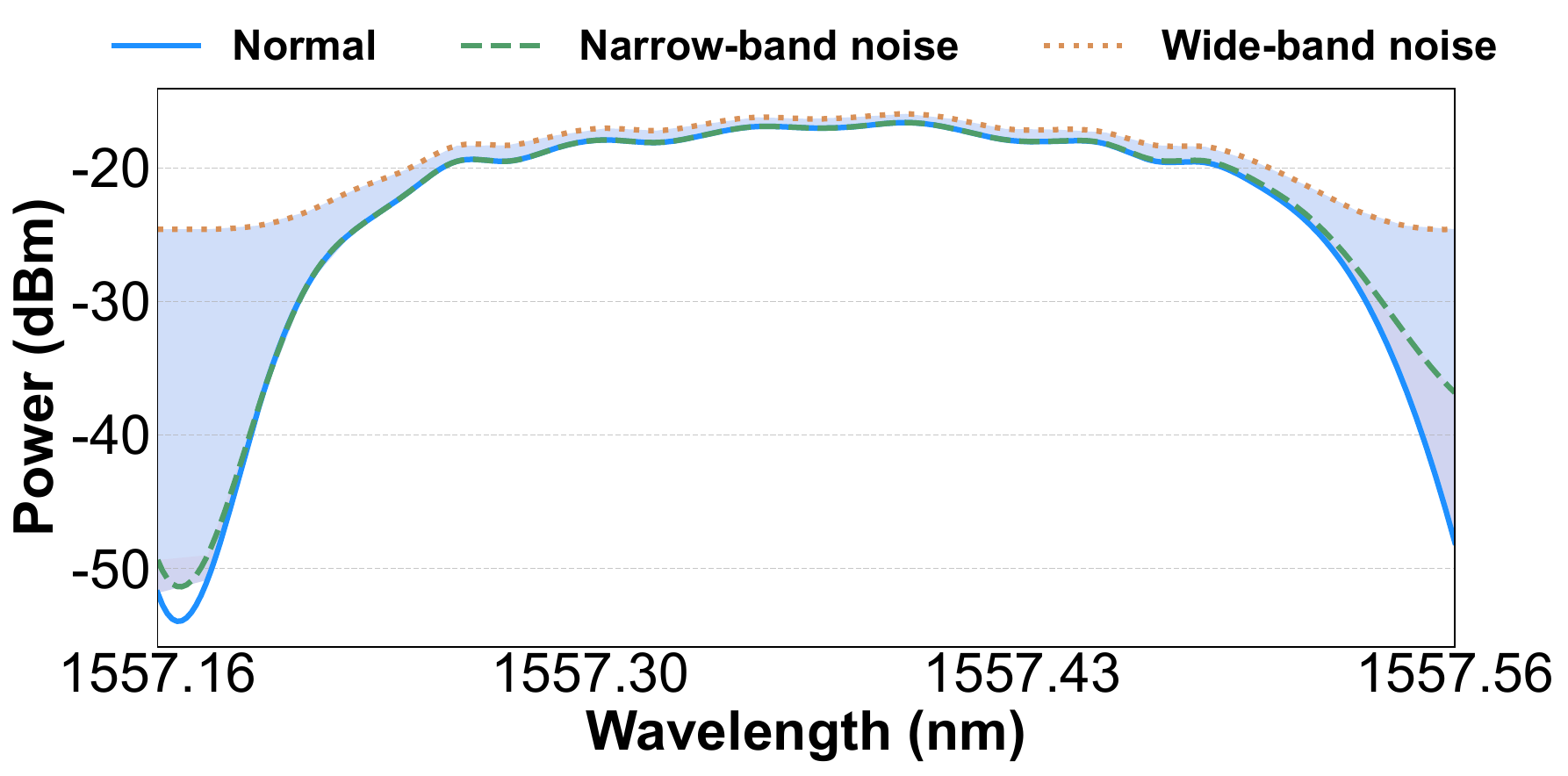}
		\label{fig:Noise_faults}
	}

	\subfigure[Spectral distortions]{
		\includegraphics[width=0.32\textwidth]{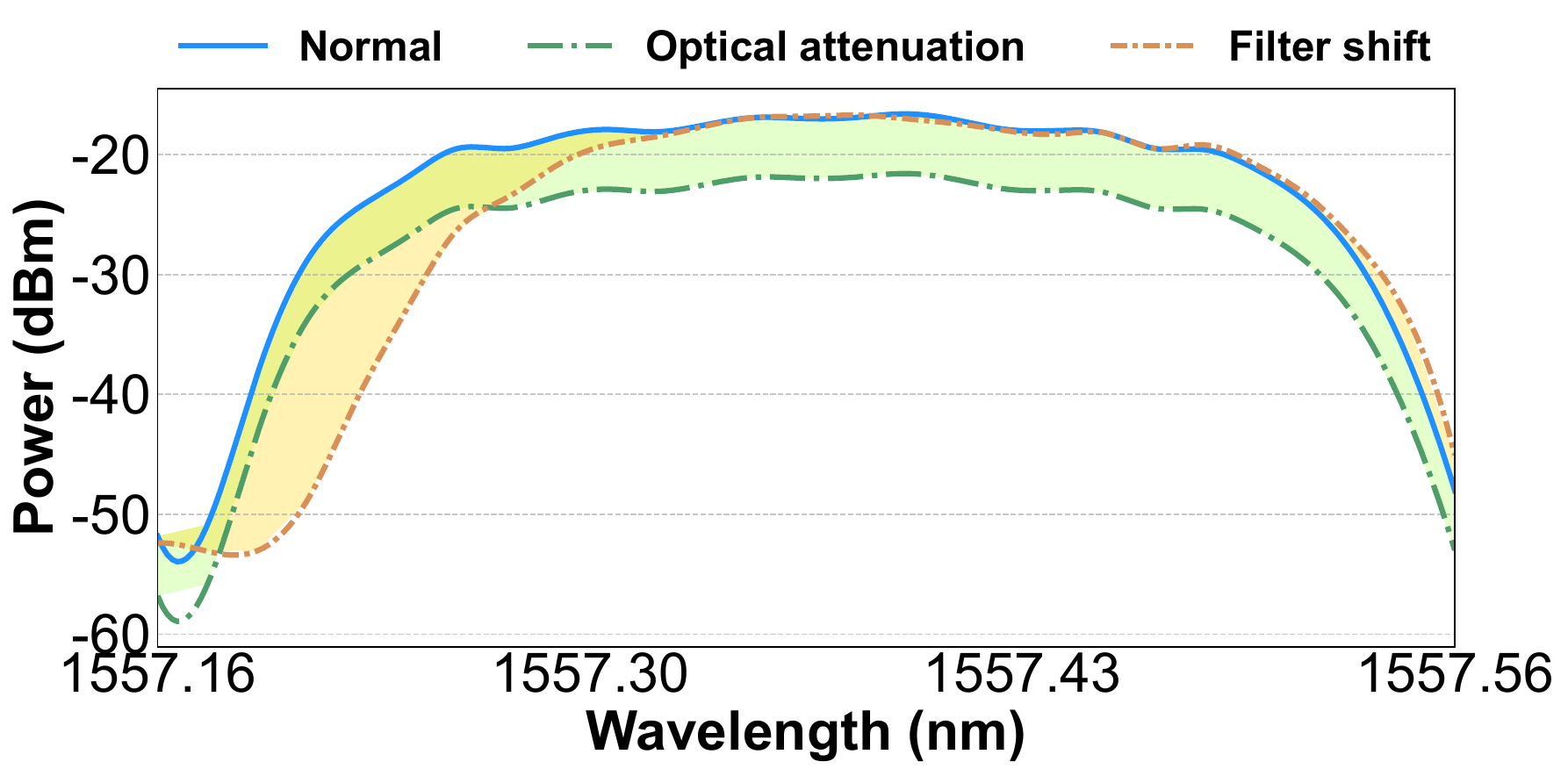}
		\label{fig:Distortion_faults}
	}
	\caption{Examples of spectral data of typical soft failures.}
	\label{fig1} 
\end{figure}

\section{Background and Motivations}

\subsection{Optical Anomaly Diagnosis} 

Optical networks suffer from both hard failures (\textit{e.g.}, fiber cuts) and soft failures (\textit{e.g.}, filter drifting)~\cite{musumeci2019tutorial, sun2022digital, vela2017ber, lechowicz2025optimizing}. Diagnosing such failures involves analyzing telemetry data collected with OPMs attached to optical nodes~\cite{chen2022automating, poggiolini2013gn, fernandez2019accurate}. As a key category of telemetry data, spectral data stands out for its ability to embody optical-layer anomalies through analyzable patterns~\cite{shariati2019learning, lun2020soft}, as exemplified in Figure~\ref{fig1}, which shows the spectra of a 50-GHz wavelength channel under various anomalies. As spectral data is normally high-dimensional (characterizing the $50$-GHz-channel with a resolution of $312.5$ MHz produces $160$-dimensional data), processing it remotely in an SDN controller or a server results in excessive latency and overheads~\cite{silva2023confidentiality, chen2022automating, chen2022cooperative,lun2023gan}, thus pushing for the shift towards distributed in-network diagnosis.

\subsection{In-Network Diagnosis with IDP} 

IDP offers the opportunity to directly embed ML models into the forwarding pipelines of programmable data plane (PDP). Nevertheless, the development of IDP has been consistently challenged by the increasing complexity of ML models. PDP operates on reconfigurable MATs~\cite{bosshart2013forwarding}, leveraging the P4-programmable protocol independent switch architecture (PISA)~\cite{bosshart2014p4}, which, by design, prioritizes packet processing throughput, introducing a few well-documented limitations. First, it is only optimized for header processing, lacking native support for general-purpose arithmetic operations. Second, its hardware resources are limited and physically partitioned over a number of stages (\textit{e.g.},  $12$ stages in a switch based on Tofino 1 ASIC~\cite{Tofino}), preventing it from storing lookup rules for high-dimensional data due to combinatorial explosion. Last but not least, its ``run-to-completion'' operation model is inherently non-iterative, and thus forces the iterative logic required by ML models to leverage multi-pass processing based on packet recirculation, incurring significant overhead in both bandwidth and latency~\cite{zheng2025p4, zhang2025quark}.

Despite the limitations above, there has been appealing progress on realizing in-network inference with IDP. Initial studies focused on DT models~\cite{xiong2019switches, busse2019pforest, zang2022p4pir, xie2023empowering, zhou2023efficient, butun2025dune}. Then, researchers also considered NNs~\cite{siracusano2022re, zhang2025quark,yan2024brain}.

The practical bottleneck in these existing NN-on-switch designs~\cite{siracusano2022re, zhang2025quark,yan2024brain} and tool-chains (\textit{e.g.}, INQ-MLT~\cite{P4_toolbox}) is not only their degraded accuracies after quantization, but also the execution restrictions of PISA. Specifically, as PDP normally cannot operate on real numbers or support matrix multiplication natively, each NN layer can only be compiled as a long sequence of integer operations, where many dot-product templates have to be assembled for the vector-matrix multiplication in it. This scenario is not suitable for directly realizing the NNs for optical anomaly diagnosis, because the input data (spectral samples) is high-dimensional, preventing a PDP pipeline from processing all the dimensions in parallel within one stage. Hence, a switch has to iterate over dimensions via stage chaining or packet recirculation, and the iteration repeats for each NN layer, amplifying latency and resource consumption as the input dimensionality and NN layer width/depth grow. Moreover, existing single-switch scenarios in this area can hardly provide the network-awareness that is critical for optical anomaly diagnosis, especially for detecting and locating complex soft failures~\cite{silva2023confidentiality, chen2022cooperative,lun2023gan}. A multi-switch scenario has been considered in~\cite{MUTA}, but it aims to mitigate per-switch resource limits by partitioning a model over multiple switches, which still can hardly gather the network-awareness for anomaly diagnosis.

\subsection{Vector Quantization}

Vector quantization (VQ) \cite{gray1984vector} is a well-known technique in signal processing for lossy data compression. It represents a vast space of high-dimensional data with a small ``codebook'' of prototype vectors. The codebook can be designed with the classic Linde-Buzo-Gray algorithm~\cite{linde2003algorithm}. This procedure allows VQ to be used for various data distributions, making it a powerful tool for discretization. VQ has a long history of success in low-bit-rate applications~\cite{gray1984vector, makhoul2005vector}, and has more recently been revived in ML through VQ-VAE~\cite{van2017neural}, where discrete codebooks are used in NN training. 

Our work is inspired by this revival. Since VQ encodes a high-dimensional vector with a compact codebook (index), it avoids the bottleneck due to per-dimension handling in existing NN-on-switch designs and aligns naturally with MAT lookups. We adopt VQ to discretize neural representations and tailor it to fit in the hardware constraints of Tofino switch. Here, the key challenge is that practical VQ encoding needs to search for the nearest neighbor~\cite{wei2000fast}, which requires the computations and memory resources that are out of the capability of Tofino switch. We resolve this issue with a software-hardware co-design that realizes VQ encoding to preprocess high-dimensional raw spectral data as integer codebook indices and architects in-network inference on Tofino switch to take the indices as inputs, thereby completely freeing the PDP hardware from complex search operations.

\subsection{GNN-based Anomaly Diagnosis}

Other than relying on a remote SDN controller, network-aware diagnosis can also be achieved with distributed GNNs. This family of NN models, including the graph convolutional network (GCN)~\cite{kipf2016GCN}, GraphSAGE~\cite{hamilton2017inductive}, and graph attention network (GAT)~\cite{velivckovic2017GAT}, all use iterative message-passing, \textit{i.e.}, they need to repeatedly aggregate information from neighbors. This leads to an intrinsic mismatch with the single-pass and non-iterative pipelines in IDP switches, making on-switch GNN tremendously challenging. The fact that optical anomaly diagnosis needs to process high-dimensional spectral data further complicates the challenge. 

In other words, the limited hardware resources in each IDP switch rule out the possibility of leveraging the straightforward implementation in~\cite{lamb2025distributed} for optical anomaly diagnosis, because it was designed to process scalar data (queue length) only.

Among GCN, GAT and GraphSAGE, GraphSAGE is the most suitable one for realizing optical anomaly diagnosis due to its relatively good hardware compatibility. Specifically, GCN requires the full global graph topology for transductive inference, and GAT computes dynamic attention scores from pairwise neighbor features. However, offloading these dynamic computations to IDP switches will consume excessive hardware resources, while the resource burden leaves less room for processing the quantized features at sufficient precision, inevitably degrading diagnostic accuracy. In contrast, GraphSAGE learns an inductive neighborhood aggregation function and employs fixed-size neighbor sampling. This deterministic mechanism aligns better with the fixed-depth pipelines and restricted hardware resources in IDP switches.

\begin{figure}[t] 
	\centering 
	\includegraphics[width=0.79\columnwidth]{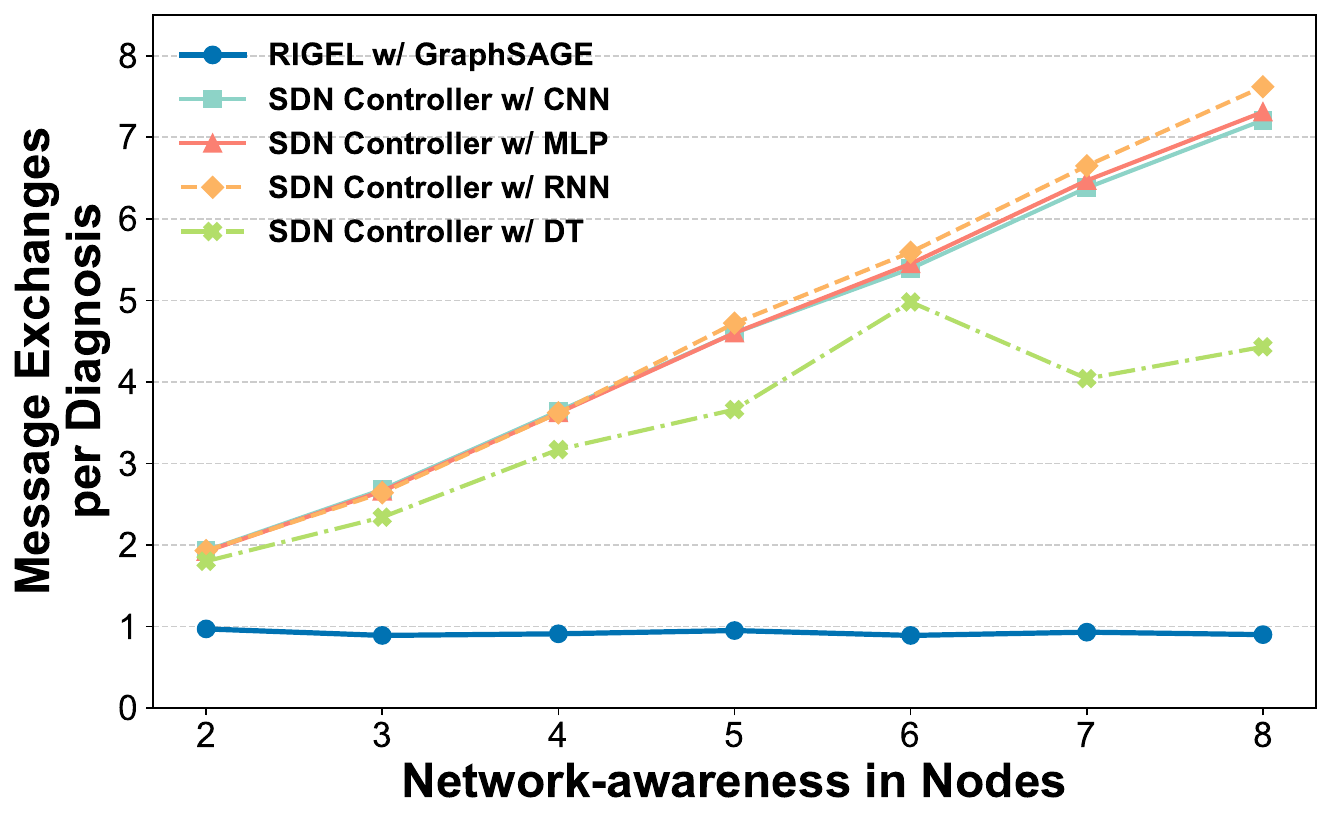}
	\caption{Communication overheads per diagnosis.}
	\label{fig2} 
\end{figure}

Therefore, our RIGEL realizes scalable and hardware-aware GraphSAGE execution on IDP switches to resolve the critical engineering challenges for optical anomaly diagnosis. Furthermore, as RIGEL confines the processing of high-dimensional spectral data entirely in IDP switches, the overheads due to message exchanges between the control and data planes are significantly reduced. 

We run simulations to quantitatively show this advantage, where the message exchanges per diagnosis are plotted in Figure~\ref{fig2}, for realizing different scales of network-awareness (in network regions containing different numbers of nodes) and comparable diagnosis accuracy. We consider five types of ML models, \textit{i.e.}, GraphSAGE (in RIGEL), multi-layer perceptron (MLP), DT, CNN and RNN, and except for RIGEL, the optical anomaly diagnosis is all realized by on-switch ML models and an SDN controller. As RIGEL can detect and locate optical anomalies, it only needs to report diagnosis results, maintaining communication overhead as constant and minimal, regardless of the scale of network-awareness. In contrast, as the ML-based centralized benchmarks are not network-aware, their message exchanges per diagnosis generally increase with the scale of network-awareness (each message exchange actually reports a much larger volume of telemetry data). Meanwhile, the average diagnosis accuracies of RIGEL, MLP, DT, CNN and RNN are 97.64\%, 96.93\%, 93.07\%, 96.97\% and 97.14\%, respectively.

\subsection{Motivations} 

Our motivations are threefold. First, the recent need for real-time and highly-efficient optical anomaly diagnosis pushes for a paradigm shift from centralized analysis to distributed in-network inference. Second, existing on-switch ML models can hardly provide the network-awareness for locating optical anomalies, and thus are not suitable for optical anomaly diagnosis. Third, while distributed on-switch GNNs are the ideal paradigm, their implementations are obstructed by the hardware restrictions that make processing high-dimensional spectral data infeasible. To this end, this work aims to design a stateful distributed in-network inference system that efficiently enables GraphSAGE-based on-switch computation for data-driven diagnostic tasks across distributed IDP switches, transforming them into a collaborative inference fabric capable of network-aware diagnosis.

\begin{figure}[t] 
	\centering 
	\includegraphics[width=0.77\columnwidth]{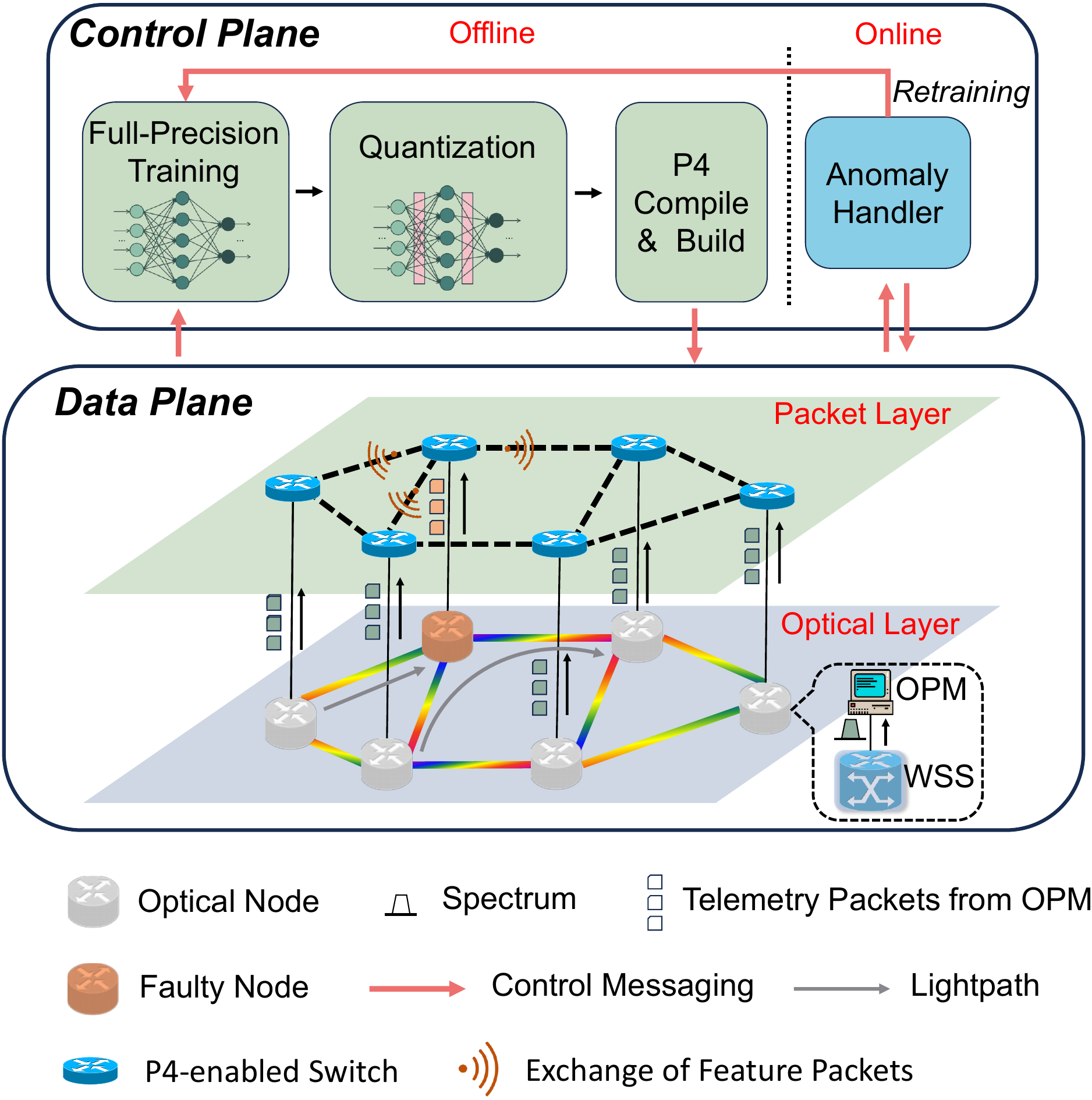}
	\caption{System overview of RIGEL.}
	\label{fig3a} 
\end{figure}

\section{Design Overview}
\label{sec:Design Overview}

We design RIGEL as a real-time optical anomaly diagnosis system that adopts the well-known packet-over-optical architecture (Figure~\ref{fig3a}). The data plane consists of a packet layer on top of an optical layer, which interconnects optical nodes with fiber links that contain in-line EDFAs for setting up lightpaths through WDM. With built-in WSS', each optical node de-multiplexes lightpaths from input fibers, terminates those that mark it as their destinations to steer to the local switch in the packet layer, and optically grooms the remaining ones with locally-generated lightpaths to send to output fibers. There is also an OPM on each optical node, to tap optical signals from its input/output fibers and perform spectrum analysis on them. The OPM samples signals at a resolution of $312.5$ MHz, and compresses and encodes the obtained spectral data as \textbf{telemetry packets} to the local switch.

Note that, as optical layer characteristics normally do not change frequently, the monitoring frequency of each OPM can just be a few times per second. Hence, the data processing rate for telemetry packets will just be in Kbps at most, which is negligible on a multi-Tbps Tofino switch.

\begin{figure}[t] 
	\centering 
	\includegraphics[width=0.69\columnwidth]{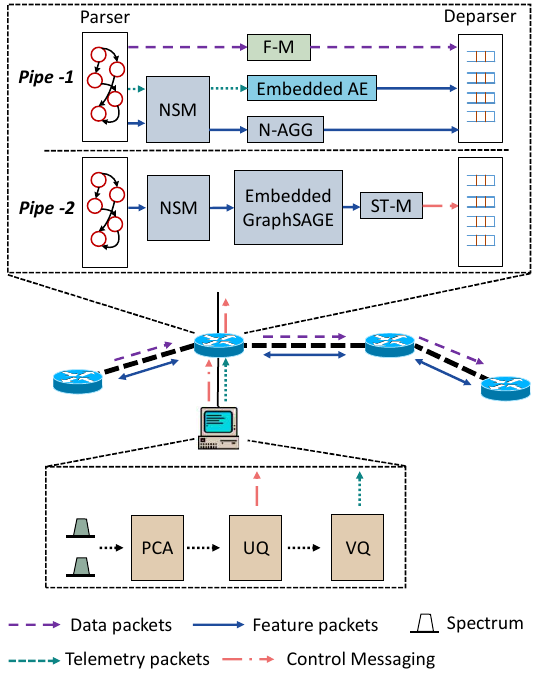}
	\caption{In-network optical diagnosis in RIGEL.}
	\label{fig3b} 
\end{figure}

\subsection{Preprocessing Raw Spectral Data in OPM}
\label{sec:preprocessing}

To achieve accurate optical anomaly diagnosis, we let each OPM collect optical spectra at its optical node and organize the data samples as follows. First, for each wavelength channel, the OPM samples the spectra of its input and output at the optical node with a resolution of $312.5$ MHz. Then, it organizes the collected data as raw spectral data samples, each of which contains the spectra of the input and output of two adjacent channels (\textit{i.e.}, each raw spectral data sample is in $640$ dimensions, if we assume a channel width of $50$ GHz). 

Apparently, the high-dimensional raw spectral data samples are not suitable for being directly processed in Tofino switch. Hence, we design the following preprocessing in the OPM to prepare the data (as shown in Figure~\ref{fig3b}). The OPM first uses a principal component analysis (PCA) module to compress raw spectral data samples into $20$ dimensions, and then leverages a per-dimension uniform quantization (UQ) to discretize the compressed real data samples to integers. Finally, we send the UQ's output to the VQ encoding module, which further maps each 20-dimensional sample to a VQ codebook index. The index is encapsulated in a telemetry packet to Tofino switch for distributed in-network inference. 

The preprocessing in OPM enables the software-hardware co-design that ensures efficient implementation of RIGEL, and provides two key benefits. First, it transforms the input to each Tofino switch from high-dimensional real samples to compact integer indices, thereby significantly reducing the complexity of our in-network inference model and making it ready to be offloaded to Tofino switch. Second, the UQ's output ($20$-dimensional integer samples) actually serves a dual-purpose: in addition to being consumed by the VQ encoding module, it can also be requested on demand by the control plane (as exemplified in Figure~\ref{fig3b}) to provide a more detailed, yet still compressed, representation of raw spectral data, enabling robust verification to resolve diagnostic ambiguities.

Note that, although the preprocessing runs on the CPU in each OPM, it is not suitable to run our GraphSAGE model on the CPU too for the following reasons. First, the CPU is just the common light-weight one in the standard configuration of OPM for spectral data collection, and thus its computing capacity is limited for supporting the GraphSAGE model with low latency. Second, running the model on a CPU can hardly avoid the delay and overhead intrinsic in hardware/software context switching. Finally, to enable graph-based neighbor aggregation, each OPM needs to expose its IP address and run network maintenance to track the optical layer topology timely, which induces both additional operational complexity and security vulnerabilities.

\subsection{In-Network Inference in Data Plane}
\label{sec:Two-stage}
Each switch in the packet layer is a P4-enabled Tofino switch, and locally connects to an optical node. The switches communicate with each other through logical links, each of which is supported by a lightpath in the optical layer. 

In addition to the task of forwarding \textbf{data packets}, we realize a two-stage in-network inference model in each switch, which uses both of its pipelines for optical anomaly diagnosis (as shown in Figure~\ref{fig3b}). Note that, each of the physical pipelines manages a dedicated subset of ports (\textit{e.g.}, \textit{Ports} 1-16 and 17-32 attach to Pipe-1 and Pipe-2, respectively). 

The OPM preprocesses high-dimensional raw spectral data samples and encodes \textbf{telemetry packets} to the embedded AE in each switch. The AE transforms the indices in telemetry packets into learned features, which are stored locally and encoded as \textbf{feature packets} to share with other switches. Meanwhile, the GraphSAGE pipeline takes in these features, which can be either locally-produced or from remote switches, to realize graph-based inference for optical anomaly diagnosis.

The operation of the two pipelines is coordinated by neighbor state machines (NSMs), which push and aggregate features from neighbor switches according to the optical layer topology. Figure~\ref{fig3b} explains RIGEL's principle of network-aware optical anomaly diagnosis. In parallel with forwarding data packets (done with the forwarding module (F-M)), Pipe-1 processes telemetry packets from the local OPM via NSM and embedded AE to get compact feature vectors, which are stored in registers and disseminated as feature packets to neighbor switches. Meanwhile, upon receiving a feature packet, the NSM buffers the features in it, and when features from a preset neighbor set have been collected, they are aggregated by the neighbor aggregation module (N-AGG).

The aggregated features are then processed by Pipe-2 for graph-based in-network inference. Specifically, the embedded GraphSAGE consumes the aggregated features (both local and remotely-generated ones) to detect and locate anomalies in the optical layer, and its results are fed to the self-test module (ST-M), which determines fault causality. If the ST-M finds that the local optical node encounters an anomaly, it reports the category and location of the anomaly to the control plane. Hence, with the neighbor aggregation of GraphSAGE, we exploit the fact that an optical anomaly normally perturbs not only the optical spectra at the faulty node but also those at its downstream nodes. Then, by incorporating the features from both local and neighbor nodes into inference, RIGEL can detect and locate anomalies that might be hard to track down from a single point of view, confining the tasks of accurate optical diagnosis entirely to the data plane. To this end, the data plane only needs to report the root cause and location of each anomaly to the control plane, ensuring efficiency and scalability by minimizing control messaging.

\subsection{Control Plane Design}
\label{sec:control_plane}

To offload the entire graph-based in-network inference onto Tofino switch, we design offline and online components in the control plane (Figure~\ref{fig3a}). The offline component first trains a full-precision model on spectral data, then uses a multi-stage feature discretization pipeline to convert it into a resource-efficient, integer-based model, and finally compiles the model into P4 artifacts for being deployed, where the learned parameters are encoded as MAT entries. 

In runtime, the control plane first uses P4Runtime to instantiate the program and populates related MATs on target switches. The online component (\textit{i.e.}, the anomaly handler) is invoked when an anomaly is detected and located, and it applies proper adjustment(s) to address the anomaly. The anomaly handler can also realize collaborative diagnosis. Specifically, when the anomaly diagnosis encounters any ambiguity or a reported anomaly needs to be verified, it requests more detailed telemetry data (the 20-dimensional integer samples from UQ) from the related OPM(s) for a more precise diagnosis by leveraging its global network view.

\subsection{Model Adaptability and Maintenance}
\label{sec:maintenance}

Finally, a practical issue to address for deploying RIGEL in real-world optical networks is how to perform maintenance to ensure model adaptivity. In the following, we discuss the schemes for realizing life-cycle maintenance of RIGEL.

\noindent\textbf{Topology Adaptation.} As the GraphSAGE model in RIGEL learns optical anomaly propagation signatures rather than memorizing a specific optical layer topology, relatively good generalization over topologies can be achieved, meaning that the model can easily adapt to a new topology after routine network expansion or minor topology changes without requiring retraining. We will verify this feature in \S\ref{subsec:New Topologies and Novel Faults}.

\noindent\textbf{Handling Unseen Anomalies.} When previously unseen anomalies occur, misclassifications will happen according to the operation principle of the GraphSAGE model. Fortunately, the incidents can be easily detected by monitoring the cases of multiple root-cause reporting (\textit{multi-root reporting}). This is because when an unseen anomaly occurs, its feature vector typically falls outside the established decision boundaries, which will make the GraphSAGE model misclassify it and trigger diverging root-cause alerts across multiple optical nodes along the lightpath. Therefore, when the number of multi-root reportings exceeds a preset threshold, the control plane will instruct the related nodes to report raw spectral data, expand the training set, and perform offline retraining. After the retraining, it deploys the updated VQ codebooks and classifier weights to Tofino switches as new MAT entries via P4Runtime, without interrupting regular packet forwarding on the switches. We experimentally demonstrate this feature and evaluate its performance in \S\ref{subsec:New Topologies and Novel Faults}.

\noindent\textbf{Regular Model Maintenance.} In a real-world optical network, the operations of network elements can drift over time (in months or even longer). Hence, it is recommended to conduct regular model maintenance, which retrains the offline model with new spectral data and recalibrates the quantization pipelines, making the model up-to-date. The retraining is performed in parallel with normal network operations and does not take excessive time ($\sim$50 minutes for a 6-node topology, and $\sim$1 hour for a larger topology with 14 nodes).

\begin{figure}[t] 
	\centering 
	\includegraphics[width=0.89\columnwidth]{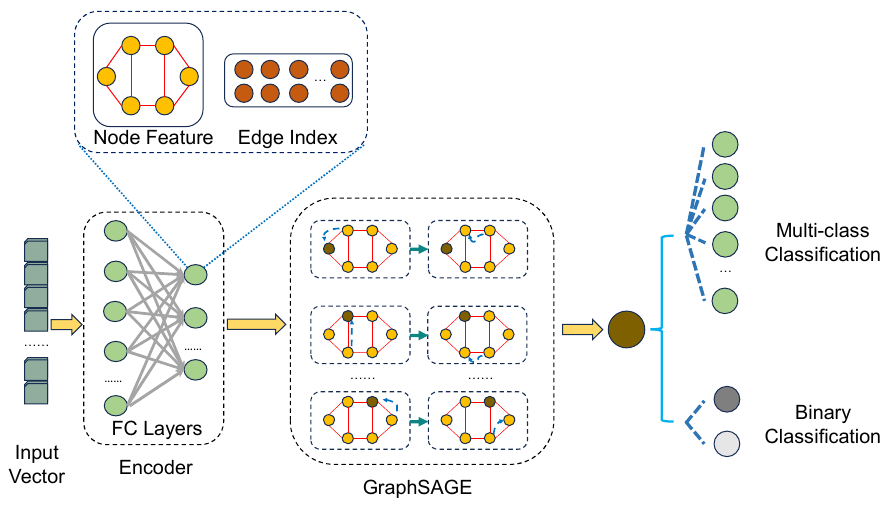}
	\caption{Architecture of full-precision AE-GraphSAGE.}
	\label{fig4} 
\end{figure} 

\section{Training of Full-Precision Model}
\label{sec:Baseline Model}

The core innovation of RIGEL lies in the offline part of its control plane, which provides a hardware-aware compiling pipeline to transform the full-precision AE-GraphSAGE model (Figure~\ref{fig4}) into a switch-deployable program. The full-precision model is in 32-bit floating-point format (FP32), and combines an AE and a GraphSAGE. The AE uses a two-layer fully-connected (FC) encoder to encode high-dimensional spectral data (the \textit{input vector}) into a compact latent representation (the initial \textit{node feature}). This reduces data dimensionality while safeguarding critical anomaly signatures. The GraphSAGE then performs inductive learning, by taking the node feature and an \textit{edge index} as inputs. Its two SAGEConv layers realize two-step aggregation. First, each node randomly samples a neighbor and uses its feature vector to create an updated representation. Second, it repeats this process, to sample another neighbor and aggregate its updated representation with that from the first step. Through this process, each node's final embedding effectively captures the information about its 2-hop neighborhood. The embedding is then fed into a final classifier to produce the output: either a binary classification for root-cause location or a multi-class classification for anomaly type identification.

The distributed mechanism of RIGEL helps to theoretically decouple the complexity of in-network inference from the number of nodes in the optical layer topology ($N$). Specifically, the graph-based neighbor aggregation of the GraphSAGE model restricts the time complexity on each switch to $O(S_1 + S_1 \cdot S_2)$, where $S_k$ denotes the number of neighbors involved in the state aggregation at hop $k$. Note that, previous studies (\textit{e.g.}, those in~\cite{kipf2016GCN, hamilton2017inductive}) have already suggested that a shallow GNN with 2 layers is often sufficient for graph learning, while applying deeper aggregation could lead to ``over-smoothing''~\cite{li2018deeper} and thus degrade the performance of graph learning. Therefore, we restrict the depth of neighbor aggregation in the GraphSAGE model to 2 hops, to maintain short diagnosis latency and realize accurate diagnosis.

We train the full-precision baseline model with labeled telemetry data collected in a real-world lab testbed (with the 6-node topology in Figure~\ref{fig3a}). Each sample contains 20-dimensional data that is compressed from the raw 640-dimensional spectral data denoting the input and output of two $50$-GHz wavelength channels, and an anomaly type. We consider $8$ types of anomalies: filter drifting in $4$ severities ($\{\pm 12.5, \pm 25\}$ GHz), abnormal power loss, broadband noise insertion, and 2 types of narrow-band noise insertion ($12.5$ GHz noise insertion at two channel edges). Then, including the normal case, each data sample is labeled as one of $9$ types. We run automatic scripts in the testbed to collect $\sim$$234,000$ samples. To reflect realistic network conditions, the dataset exhibits a natural class imbalance, including $\sim$$192,000$ normal samples ($85.19\%$) and $\sim$$33,000$ anomalous samples ($14.81\%$), with $\sim$$4,000$ samples for each of the $8$ anomaly types. We divide the dataset into training, validation, and testing sets according to a split of $0.6:0.2:0.2$. The baseline model is trained for $500$ epochs with a batch size of $128$, and after training, its accuracies on anomaly detection and location both exceed $99\%$ on the testing set.

\section{Multi-Stage Feature Discretization}
\label{sec:Multi-stage}

With the trained full-precision AE-GraphSAGE model, we design a multi-stage feature discretization pipeline to convert it into an integer-based, MAT-friendly model for on-switch implementation. The pipeline includes a UQ followed by a VQ. The UQ maps each dimension of an FP32 input vector to an integer on a grid with a calibrated step and zero point, obtaining an integer UQ vector (detailed procedure in Appendix \S\ref{sec:UQ}). Then, VQ takes the integer vector and finds its closest match within a learned VQ codebook (see \S\ref{sec:VQ}). The index of the best-matching entry (\textit{i.e.}, the VQ index) serves as the final compressed result fed to subsequent processing on Tofino switch. This VQ mechanism is applied both before the AE to compress its input and before each aggregation in GraphSAGE to compress input feature vectors for MAT-driven fusion. Finally, the resulting artifacts of the discretized model, including the VQ codebooks (for AE and GraphSAGE) and small fusion/latent lookup tables, are compiled into P4 MATs and registers, forming an integer-only and table-driven datapath for optical anomaly diagnosis.

\begin{figure}[t] 
	\centering 
	\includegraphics[width=0.88\columnwidth]{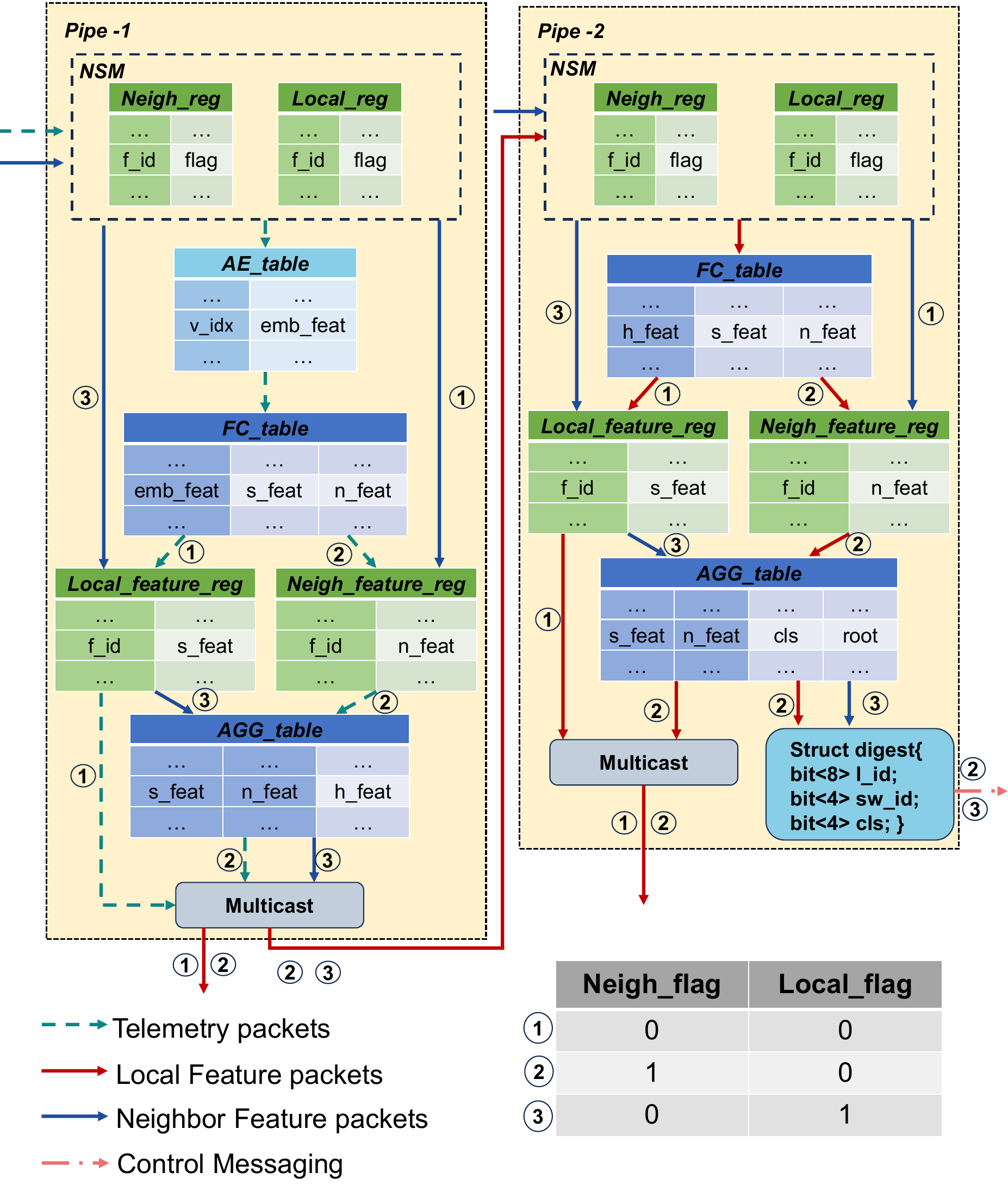}
	\caption{Data plane architecture of RIGEL.}
	\label{fig:RIGEL_architecture} 
\end{figure}

\section{Data Plane Implementation}

Figure~\ref{fig:RIGEL_architecture} shows the implementation of RIGEL in Tofino switch, which uses both of its physical pipelines (Pipe-1 and Pipe-2). Each pipeline is coordinated by an NSM, which handles the asynchronous arrivals of local telemetry packets and feature packets from neighbors. Specifically, the NSM utilizes paired register arrays (\texttt{Local\_reg} and \texttt{Neigh\_reg}), in which each entry is indexed by a lightpath feature ID (\texttt{f\_id}) and maintains a 1-bit \texttt{flag}, to record the state of the corresponding feature processing. Here, \texttt{f\_id} is a global ID to index the spectral feature of a lightpath, \textit{i.e.}, it uses $16$ bits to point to the unique combination of a lightpath and the network time of its spectrum measurement. As shown at the right corner of Figure~\ref{fig:RIGEL_architecture}, for each lightpath feature, its pair of \texttt{Local\_reg} and \texttt{Neigh\_reg} denotes three pending states, where $1$ means the corresponding packet has been received, and $0$ otherwise. Then, based on these states, an NSM coordinates its pipeline to process an incoming packet as follows:

\noindent\textbf{State \ding{172} (Initial arrival \& wait):} If both \texttt{Local\_flag=0} and \texttt{Neigh\_flag=0}, the packet is the first one for the specific lightpath feature. If the packet is local, Pipe-1 for a telemetry packet or Pipe-2 for a local feature packet respectively extracts the self-feature (\texttt{s\_feat}) and neighbor-feature (\texttt{n\_feat}) from it, via the corresponding tables (\texttt{AE\_table} and \texttt{FC\_table} in Pipe-1 or \texttt{FC\_table} in Pipe-2). It caches \texttt{s\_feat} in \texttt{Local\_feature\_reg}, multicasts \texttt{n\_feat} to neighbors, updates \texttt{Local\_flag} to 1, and terminates to wait. Otherwise, if the packet is from a neighbor, NSM directly caches its feature (\texttt{n\_feat}) in \texttt{Neigh\_feature\_reg}, updates \texttt{Neigh\_flag} to 1, and terminates to wait.

\noindent\textbf{State \ding{173} (Local-triggered aggregation):} If NSM only detects \texttt{Neigh\_flag=1}, the incoming packet is local and its corresponding neighbor feature packet has already come in. It gets \texttt{s\_feat} and \texttt{n\_feat} and multicasts \texttt{n\_feat} as in State \ding{172}, and triggers feature aggregation with \texttt{AGG\_table}. Then, Pipe-1 sends the intermediate feature (\texttt{h\_feat}) from \texttt{AGG\_table} to Pipe-2, where the \texttt{AGG\_table} yields the anomaly class (\texttt{cls}) and location (\texttt{root}). If \texttt{root=1}, a digest is generated to alert the control plane, where \texttt{l\_id} and \texttt{sw\_id} index the lightpath and optical node, respectively (Figure~\ref{fig:RIGEL_architecture}). 

\noindent\textbf{State \ding{174} (Neighbor-triggered aggregation):} If NSM only detects \texttt{Local\_flag=1}, the new packet is a neighbor feature one and its corresponding local packet has been received. It retrieves the cached \texttt{s\_feat} and triggers feature aggregation. Then, similar to the processing for State \ding{173}, Pipe-1 forwards its aggregated result to Pipe-2, while Pipe-2 performs the final diagnosis and conditionally uploads the digest if \texttt{root=1}.
    
\noindent\textbf{Drop:} To protect the switch from resource exhaustion, any feature packet with an invalid \texttt{f\_id}, an unmatched neighbor ID, or belonging to an expired diagnostic session is dropped. This essential fallback ensures the registers and MATs are not blocked by erroneous or delayed traffic.

\section{Experimental Evaluations}

\subsection{Experimental Setup}

\noindent\textbf{Network Testbed.} We prototype and test RIGEL in a realistic packet-over-optical network testbed. Each optical node is built on Finisar 1$\times$9 bandwidth-variable WSS' (BV-WSS') (bandwidth granularity at $12.5$ GHz), and the bandwidth-variable transponders (BVTs) on a Juniper BTI-7800 platform, where each BVT uses a channel width of $50$ GHz to achieve the data-rate of $100$ Gbps with QPSK modulation ($31.2$ Gbaud with forward-error correction). The optical nodes are interconnected by standard single-mode fiber links, each of which contains an in-line EDFA, according to a topology with $[6,14]$ nodes, where the 6-node topology is in Figure~\ref{fig3a} while the 14-node one is NSFNET topology \cite{zhu2013nsfnet} (Except for those in \S\ref{subsec:New Topologies and Novel Faults}, our experiments all use the 6-node topology). For simplicity, we emulate the transmission loss of each fiber link with a variable optical attenuator (VOA).

Each OPM is implemented with a Finisar high-resolution optical channel monitor (OCM) and a mini-PC board with Intel CPU, which are both easily-accessible off-the-shelf products. The OCM can scan the whole spectrum of C-band with a resolution of $312.5$ MHz within two seconds and the mini-PC board realizes the pipelined process (PCA$\to$UQ$\to$VQ, shown in Figure \ref{fig3b}) to compress and encode raw spectral data from the OCM as telemetry packets. 

Each packet switch in the testbed is a PDP switch based on Tofino 1 ASIC.

\noindent\textbf{Anomaly Scenarios.} We consider $8$ categories of soft failures: 1) drifting of the center frequency of a BV-WSS, whose severities are characterized as $\{\pm12.5, \pm25\}$ GHz, denoted as anomaly \textit{Classes} 1-4, respectively, 2) abnormal power loss on a fiber link (\textit{Class} 5), 3) broadband noise insertion (\textit{Class} 6), realized by tuning an EDFA, and 4) two types of narrow-band noise insertion (\textit{Classes} 7-8), generated by filtering the amplified spontaneous emission (ASE) noise from an EDFA to $12.5$ GHz and respectively inserting it at two edges of a wavelength channel. Note that, all the anomalies only induce mild degradation but will not interrupt any lightpath.

\noindent\textbf{Model and Metrics.} The full-precision baseline model contains a 2-layer fully-connected AE (with a hidden width of 256) where the encoder's output is 10-dimensional, followed by a 2-layer GraphSAGE whose sampled neighbors per aggregation are set to 1. All the quantized variants for being offloaded on Tofino switch strictly share this architecture and are trained with the multi-step procedure in \textit{Algorithm}~\ref{alg:vq-multiphase} (\S\ref{sec:UQ}). The metrics considered by our evaluations are the \textbf{accuracy of anomaly classification} (AccCls) and its corresponding \textbf{F1-score} (F1Cls), as well as the \textbf{accuracy of root-cause location} (AccLoc) and its \textbf{F1-score} (F1Loc).

\begin{table}[h]
\centering\footnotesize
\caption{\centering End-to-End Diagnostic Performance.}
\label{tab:e2e_performance}
\begin{tabular}{c c c c}
\toprule
\textbf{AccCls (\%)} & \textbf{F1Cls (\%)} & \textbf{AccLoc (\%)} & \textbf{F1Loc (\%)} \\ \midrule
99.47 & 98.11 & 99.54 & 98.41 \\ \bottomrule
\end{tabular}
\end{table}

\subsection{End-to-End Performance}
\label{subsec:e2e_performance}

We first evaluate the end-to-end performance of RIGEL to verify its effectiveness for optical anomaly diagnosis.

\begin{figure}[t] 
	\centering 
	\subfigure[Anomaly classification]{
		\includegraphics[width=0.47\columnwidth]{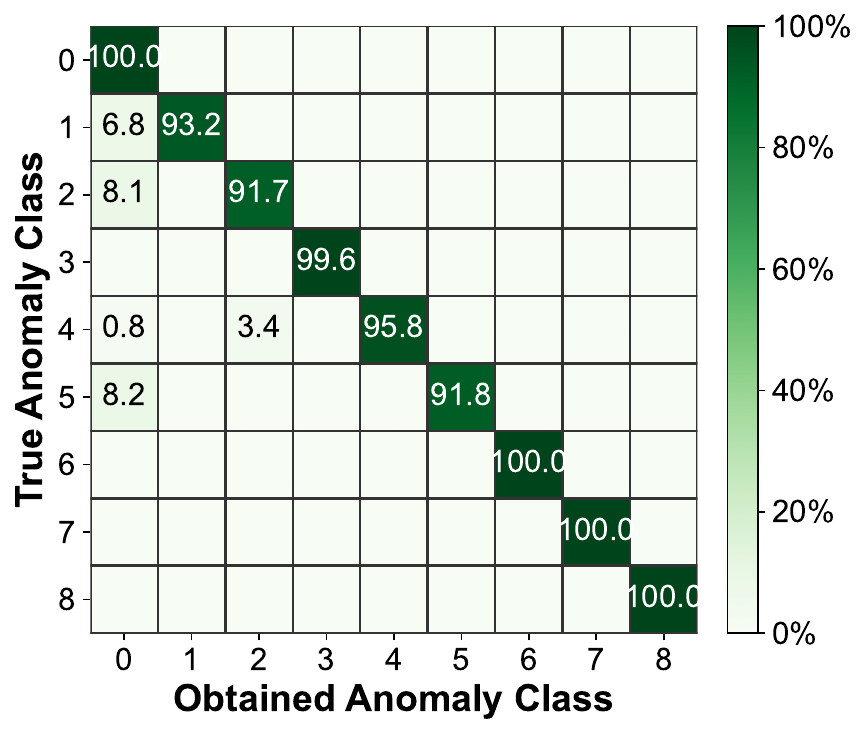}
		\label{fig:ex_tofino_a}
	}
	\subfigure[Root-cause location]{
		\includegraphics[width=0.47\columnwidth]{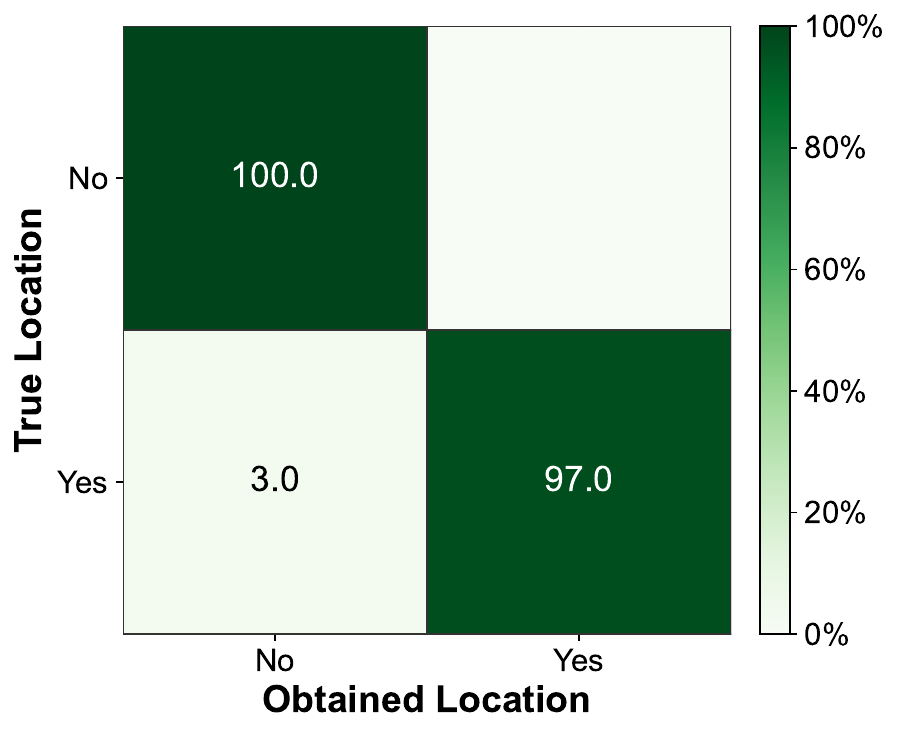}
		\label{fig:ex_tofino_b}
	}
	\caption{Confusion matrices by RIGEL for diagnosis tasks.}
	\label{ex_tofino} 
\end{figure}

\noindent\textbf{Diagnostic Accuracy.} Table~\ref{tab:e2e_performance} summarizes RIGEL's end-to-end performance on the tasks of anomaly classification and root-cause location. We can see that all the accuracy metrics are above 98\%, demonstrating exceptional diagnosis accuracy. Figure~\ref{ex_tofino} shows the confusion matrices of the two tasks to provide a detailed per-class breakdown, confirming the stable diagnosis achieved by RIGEL for each class.

\noindent\textbf{In-Switch Inference Latency.} Beyond high accuracy, a more important benefit of the in-switch processing in RIGEL is that it can effectively accelerate optical anomaly diagnosis. To explicitly quantify this latency advantage, we conduct a microbenchmark experiment to compare the running time of data processing in one GraphSAGE aggregation layer, on a Tofino switch and a CPU server with dual Intel Xeon Silver 4210 CPUs at 2.20 GHz and 128 GB memory. To ensure the comparison focuses solely on per-node processing in each system, we deliberately exclude the latency due to inter-node communication. The results indicate that processing a spectral data sample takes $\sim$10~ms on the CPU server, while the same data processing can be accomplished in $\sim$800~ns with exact MATs on the Tofino switch.

\noindent\textbf{Hardware Resource Utilization.} Table~\ref{tab:resource_usage} lists the hardware footprint of RIGEL on Tofino 1 ASIC (compiled by Intel P4 Studio SDE). We can see that by explicitly decoupling feature computation and avoiding packet recirculation, RIGEL utilizes the hardware resources efficiently. Specifically, it only uses 6 of the 12 physical stages in both pipelines, and its memory footprint is also modest, using only 3.8\% and 2.7\% of the SRAM in Pipe-1 and Pipe-2, respectively. As it does not rely on ternary rules, RIGEL does not use any TCAM. Beyond storage, RIGEL imposes low pressure on the internal lookup datapath: the exact-match search/result buses are below 9\% in both pipelines, and the action data bus is at most 2.6\%. In all, these results verify that RIGEL only takes a small portion of the hardware resources in Tofino ASIC, ensuring hitless co-existence with other in-switch functions.

\begin{table}[h]
\centering\footnotesize
\caption{\centering Hardware Resource Utilization of RIGEL.}
\label{tab:resource_usage}
\begin{tabular}{lcc}
\toprule
\textbf{Resource Type} & \textbf{Pipe-1} & \textbf{Pipe-2} \\ \midrule
Stages               & 6/12   & 6/12 \\ 
SRAM                 & 3.8\%  & 2.7\%  \\
TCAM                 & 0.0\%  & 0.0\%  \\
Logical Tables       & 17.2\%  & 14.6\% \\
Exact Match Search Bus        & 8.9\% & 7.8\% \\
Exact Match Result Bus        & 8.9\% & 8.3\% \\
Action Data Bus Bytes         & 2.6\%  & 1.4\%  \\ \bottomrule
\end{tabular}
\end{table}

\begin{figure}[t] 
	\centering 
	\subfigure[Accuracy \textit{versus} codebook bit-width]{
		\includegraphics[width=0.34\textwidth]{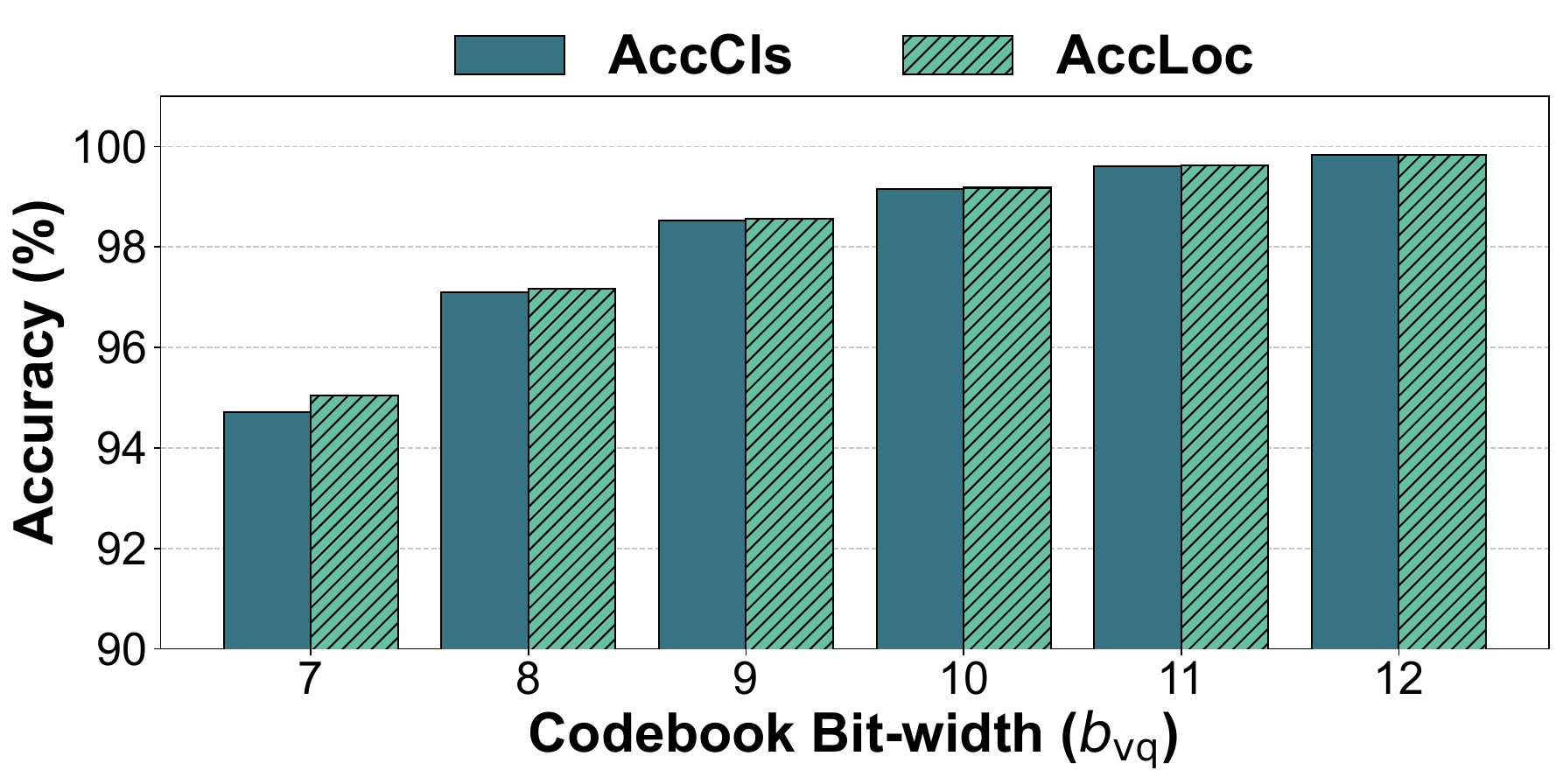}
		\label{fig:ex_1_a}
	}

	\subfigure[F1-score \textit{versus} codebook bit-width]{
		\includegraphics[width=0.34\textwidth]{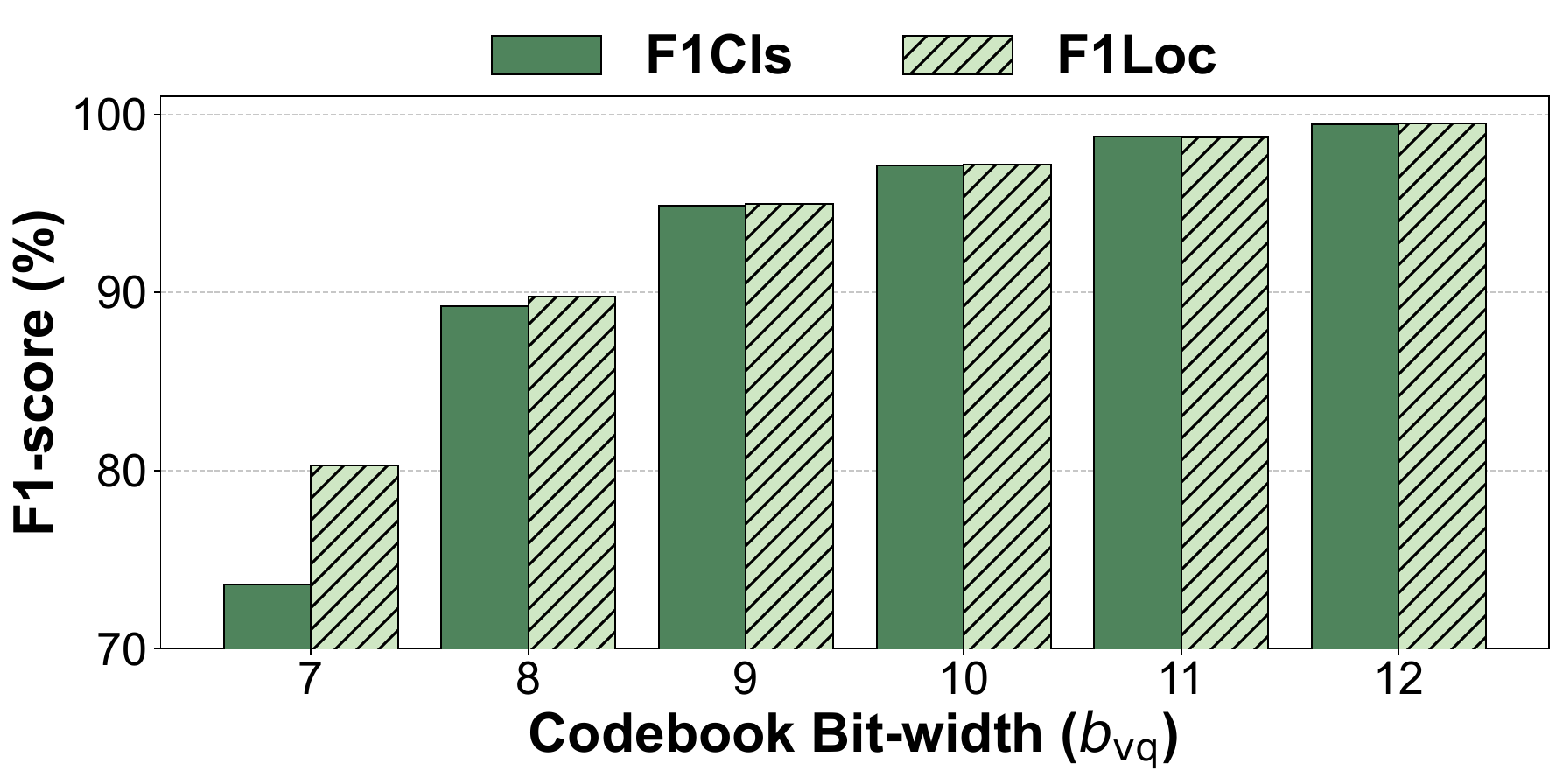}
		\label{fig:ex_1_b}
	}
	\caption{Impact of VQ codebook bit-width ($b_{\text{vq}}$).}
	\label{fig_ex_1} 
\end{figure}

\subsection{Impact of Quantization Bit-width}
We then try to understand the contribution and trade-off of each component in RIGEL. Specifically, we test the components incrementally, starting from the initial VQ-only system (only applying VQ before AE) to the full implementation of RIGEL. For benchmarking, we have verified that the accuracies (AccCls and AccLoc) and F1-scores (F1Cls and F1Loc) of the full-precision baseline are all above $99\%$.

\noindent\textbf{Impact of $b_{\text{vq}}$ on VQ-only System.}
We first evaluate the impact of the VQ with a ``VQ-only'' configuration, where the floating-point latent vectors from the PCA in an OPM are directly mapped to their nearest codewords in codebook $C_{\text{in}}$ by a VQ before being passed to the AE+GraphSAGE. Note that, as we would like to focus exclusively on the impact of the VQ before AE, the VQ applied before each aggregation in GraphSAGE ($C_{\text{agg}}$) is not used in the experiments.

Figure~\ref{fig_ex_1} indicates that increasing the bit-width of VQ codebook ($b_{\text{vq}}$) can obtain significant gains in both accuracies and F1-scores, while the gain becomes marginal after $b_{\text{vq}}$ reaches $11$. As the MAT size of VQ codebook scales exponentially with $b_{\text{vq}}$, to strike a proper balance between performance and memory efficiency, we will select $b_{\text{vq}} \in [9, 11]$ in subsequent experiments. Figure~\ref{fig_ex_1} also indicates that the system does not work well when the bit-width is $b_{\text{vq}}=7$, as the relatively low F1-scores indicate that the VQ-only configuration fails to distinguish between certain anomaly classes. 

\begin{figure}[t] 
	\centering 
	\subfigure[$b_\text{vq}=7$]{
		\includegraphics[width=0.47\columnwidth]{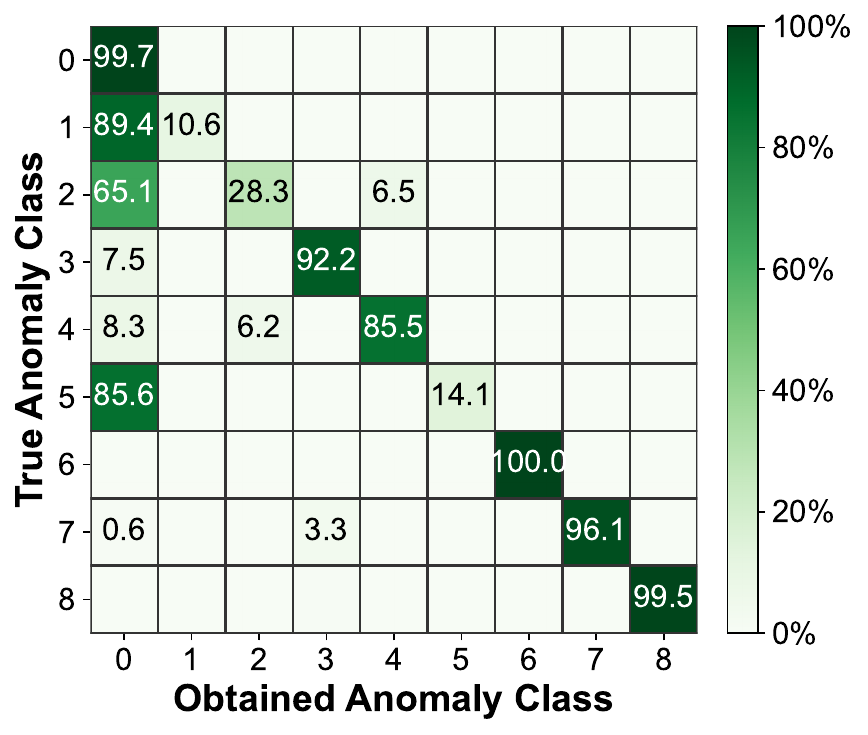}
		\label{fig:ex_2_a}
	}
	\subfigure[$b_\text{vq}=12$]{
		\includegraphics[width=0.47\columnwidth]{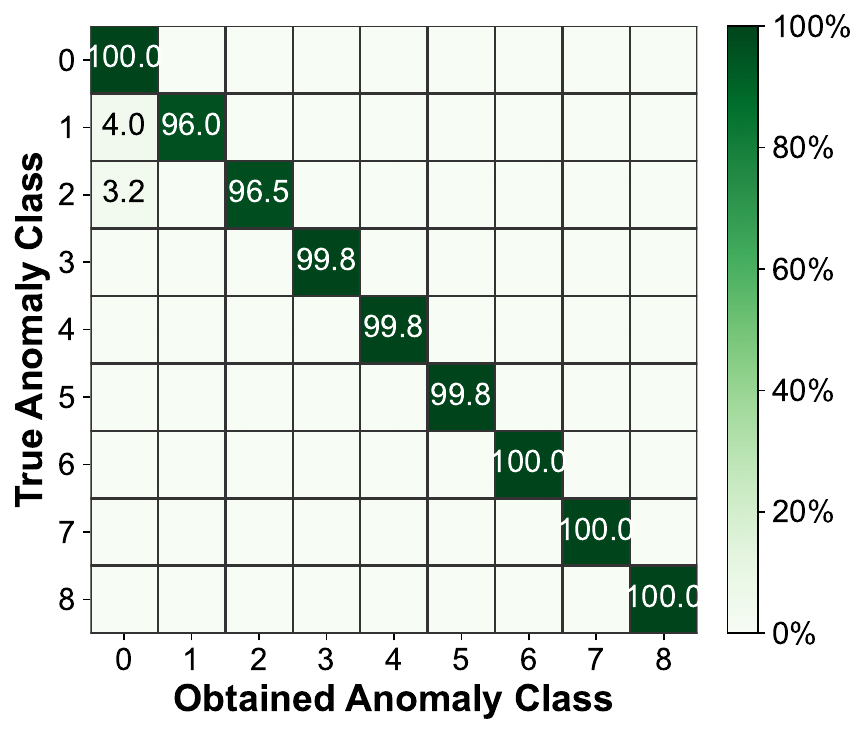}
		\label{fig:ex_2_b}
	}
	\caption{Confusion matrices for anomaly classification.}
	\label{fig_ex_2} 
\end{figure}

\begin{figure}[t]
	\centering 
        \subfigure[$b_\text{vq}=7$]{
		\includegraphics[width=0.47\columnwidth]{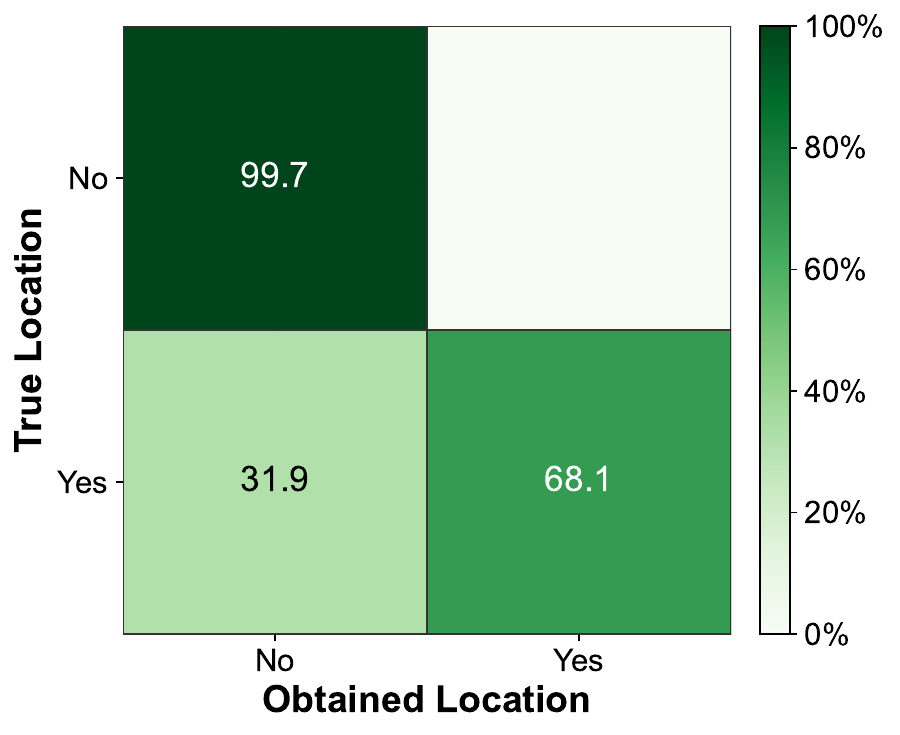}
		\label{fig:ex_3_a}
	}
        \subfigure[$b_\text{vq}=12$]{
		\includegraphics[width=0.47\columnwidth]{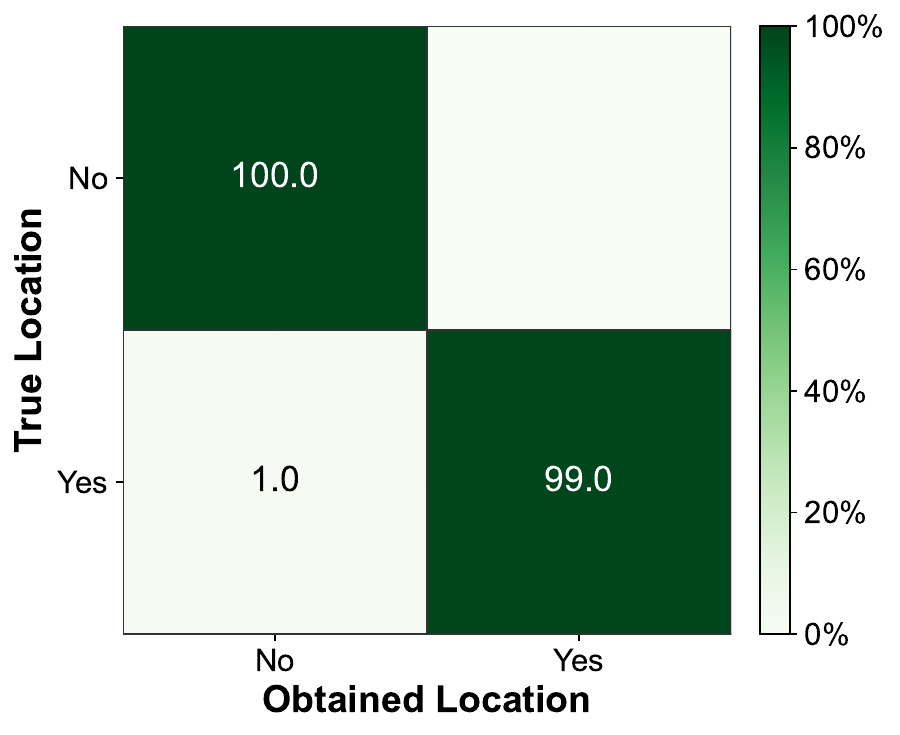}
		\label{fig:ex_3_b}
	}
        \caption{Confusion matrices for root-cause location.}
	\label{fig_ex_3} 
\end{figure}

The analysis above can be confirmed by the confusion matrix in Figure~\ref{fig:ex_2_a}, where noticeable confusions happen among anomaly \textit{Classes} 1, 2 and 5, which correspond to relatively small spectral changes by a BV-WSS drifting of $\pm12.5$ GHz (\textit{Classes} 1 and 2) and abnormal power loss (\textit{Class} 5). Therefore, with a relatively small codebook ($K_\text{in}=2^{b_{vq}}=128$), the quantization is too coarse to distinguish small spectral changes, leading to low F1-scores. In contrast, after we increase the bit-width to $b_{\text{vq}}=12$, the codebook provides enough granularity to distinguish the anomaly classes, and thus the confusion matrix in Figure~\ref{fig:ex_2_b} is almost diagonal. Similarly, for the task of root cause location, using a small codebook with $b_{\text{vq}}=7$ also leads to a high rate of missing the true location of an anomaly ($31.9\%$), as shown in Figure~\ref{fig:ex_3_a}, while the large codebook with $b_{\text{vq}}=12$ effectively eliminates the confusion (Figure~\ref{fig:ex_3_b}). 

\begin{table}[t]
	\centering\footnotesize
	\caption{\centering Performance of Full Implementation of RIGEL.}
	\label{tab:hyperparam_results}

        \begin{tabular}{
          ccc
          S[table-format=2.2]
          S[table-format=2.2]
          S[table-format=2.2]
          S[table-format=2.2]
        }

		\toprule
		\multicolumn{3}{c}{Hyper-parameters} & \multicolumn{2}{c}{Cls } & \multicolumn{2}{c}{Loc } \\
		\cmidrule(r){1-3} \cmidrule(lr){4-5} \cmidrule(l){6-7}
        
		\makecell{$b_\text{vq}$} & \makecell{$b_\text{uq}$} & \makecell{$b_\text{agg}$} & {Acc ($\%$)} & {F1 ($\%$)} & {Acc ($\%$)} & {F1 ($\%$)} \\
		\midrule

	    10 & 4 & 6 & 97.07 & 88.32 & 97.50 & 91.04 \\
		10 & 5 & 6 & 96.57 & 85.30 & 97.37 & 90.62 \\
		10 & 6 & 6 & 97.25 & 88.14 & 98.10 & 93.31 \\
		11 & 4 & 6 & 98.63 & 94.46 & 98.90 & 96.18 \\
		11 & 5 & 6 & 96.29 & 79.92 & 97.18 & 89.63 \\
		11 & 6 & 6 & 97.17 & 87.32 & 98.26 & 93.87 \\
		
		\midrule

	    10 & 4 & 7 & 97.56 & 89.74 & 98.16 & 93.55 \\
		10 & 5 & 7 & 97.89 & 92.89 & 98.24 & 93.86 \\
		10 & 6 & 7 & 96.77 & 86.30 & 97.19 & 89.83 \\
		11 & 4 & 7 & 98.78 & 94.30 & 99.33 & 97.68 \\
		11 & 5 & 7 & 99.18 & 96.83 & 99.37 & 97.83 \\
        11 & 6 & 7 & {\bfseries 99.45} & {\bfseries 98.06} & {\bfseries 99.52} & {\bfseries 98.35} \\
		
		\bottomrule
	\end{tabular}
\end{table}

\begin{figure}[t] 
	\centering 
	\subfigure[AccCls for anomaly classification]{
		\includegraphics[width=0.32\textwidth]{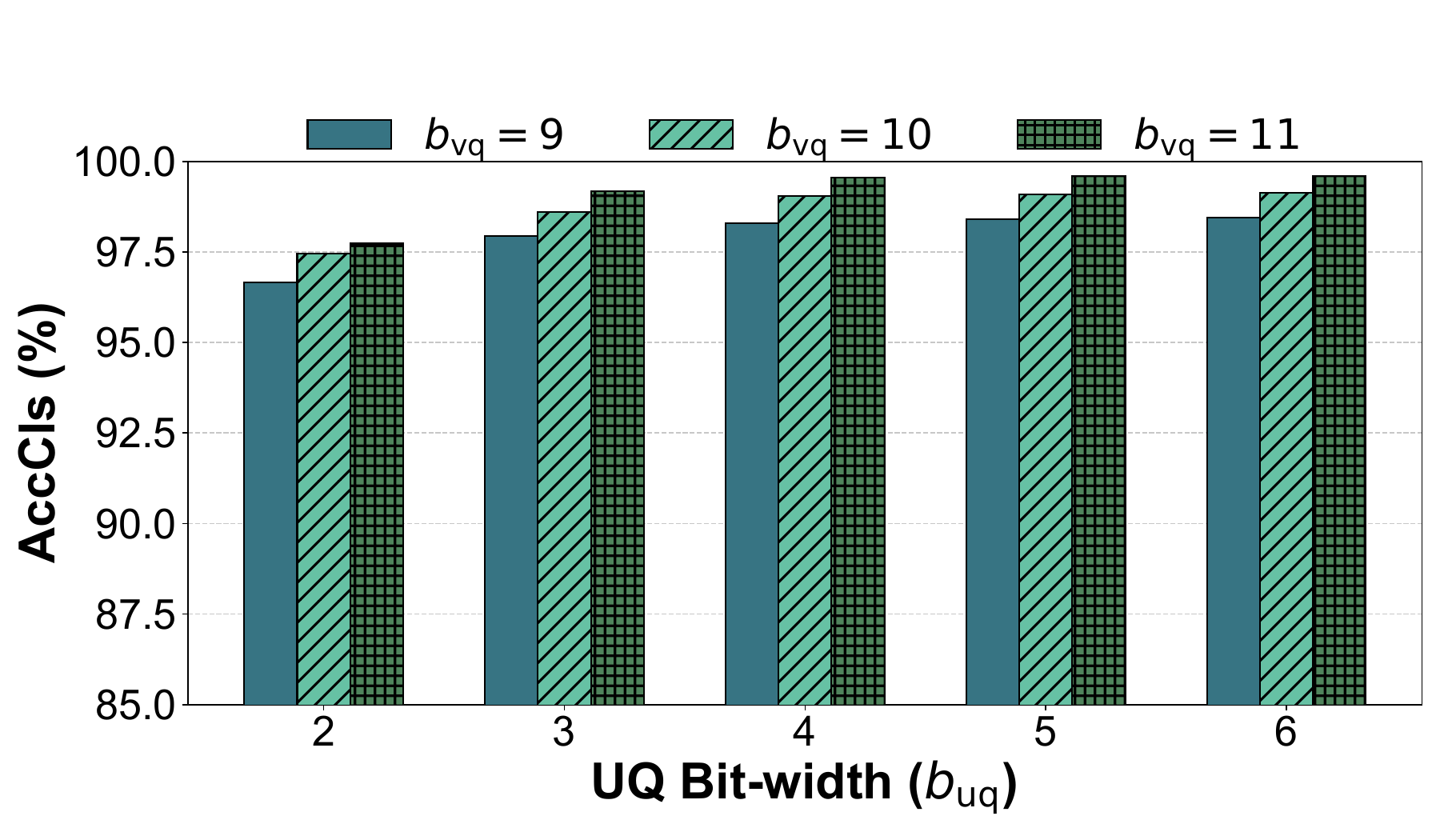}
		\label{fig:ex_4_a}
	}

	\subfigure[F1Cls for anomaly classification]{
		\includegraphics[width=0.32\textwidth]{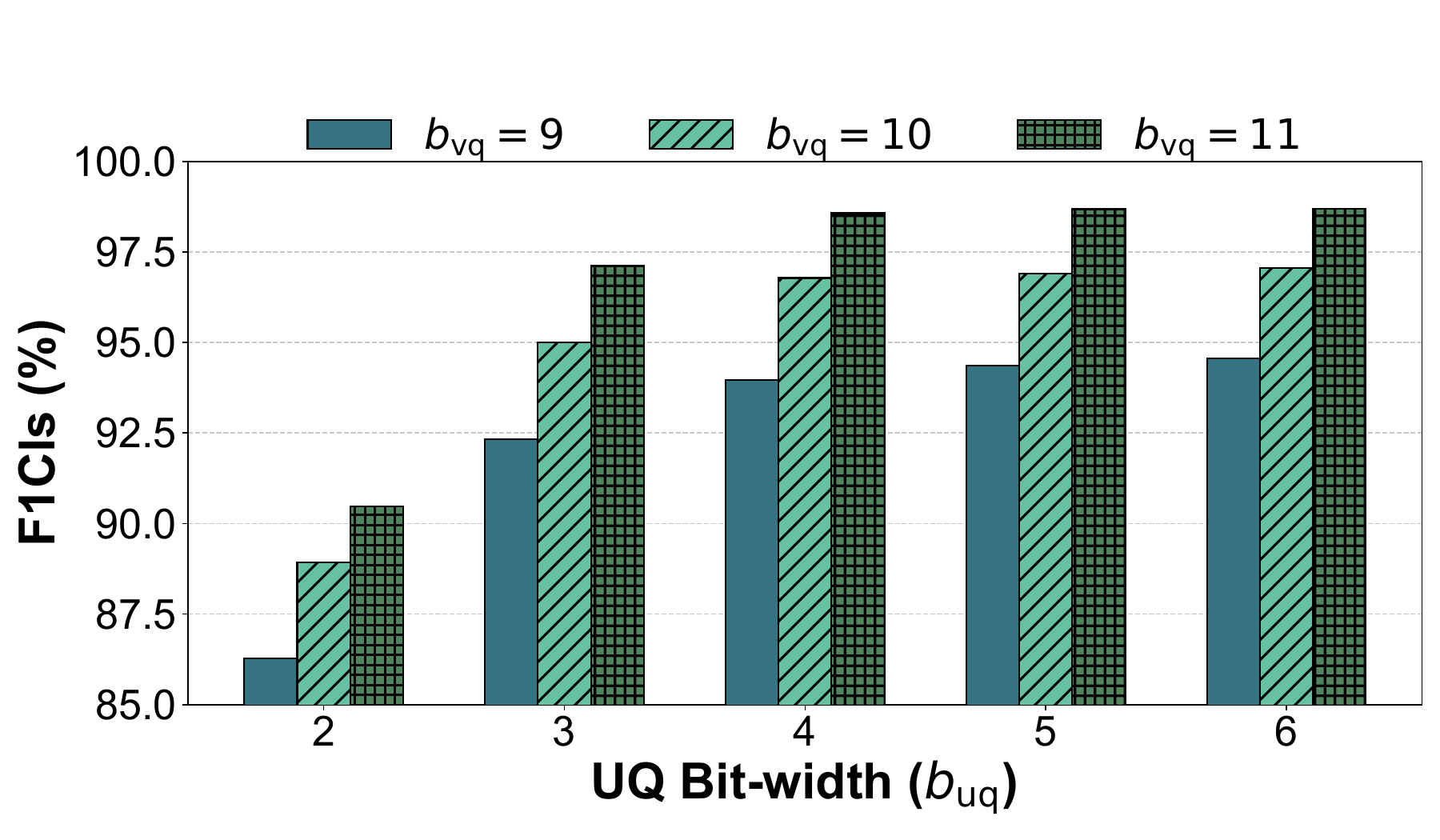}
		\label{fig:ex_4_b}
	}
	\caption{Impact of UQ codebook bit-width ($b_{\text{uq}}$) on Cls.}
	\label{fig_ex_4_1} 
\end{figure}

\begin{figure}[t]
\centering
        \subfigure[AccLoc for root-cause location]{
		\includegraphics[width=0.32\textwidth]{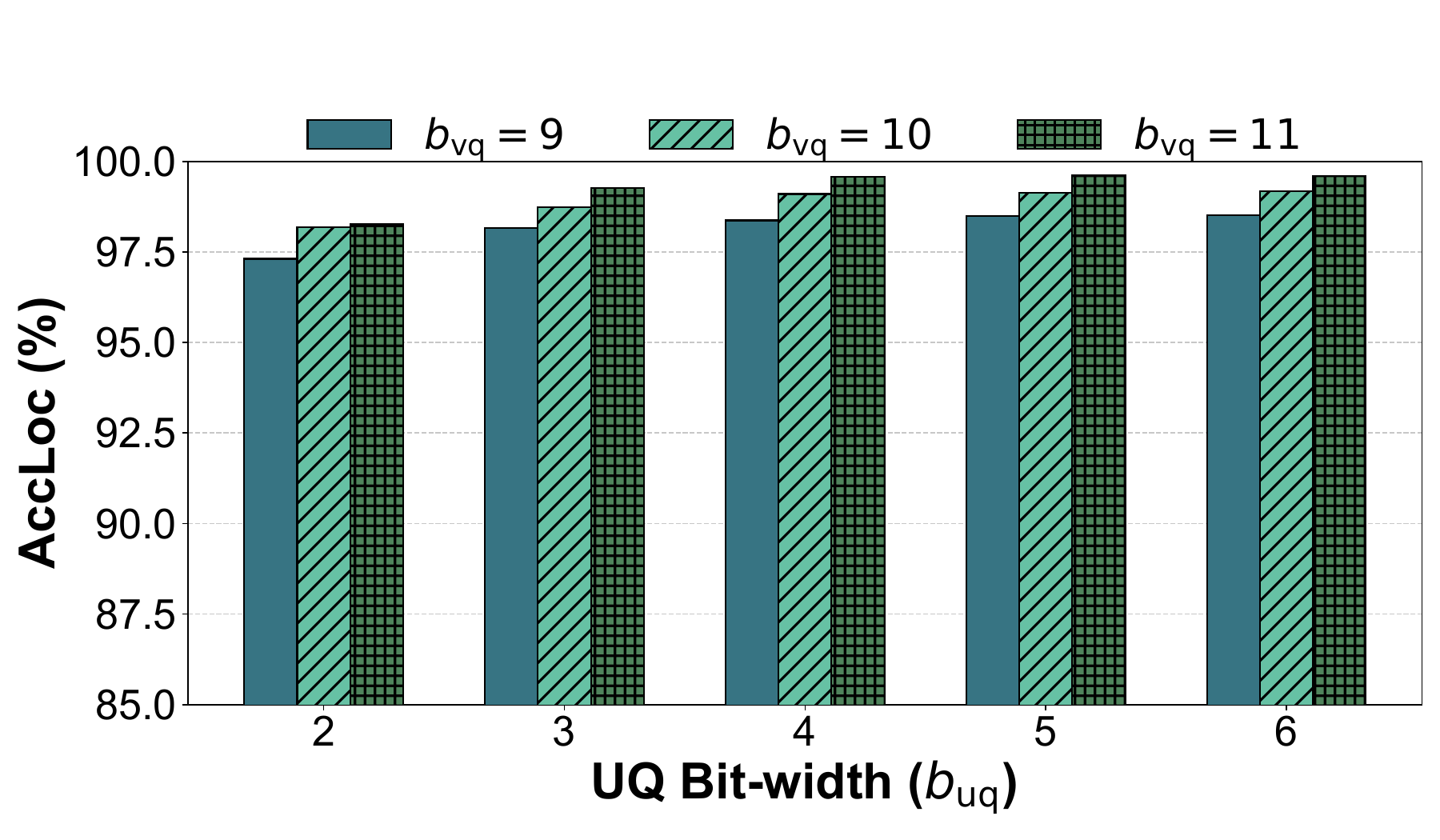}
		\label{fig:ex_4_c}
	}
        \subfigure[F1Loc for root-cause location]{
		\includegraphics[width=0.32\textwidth]{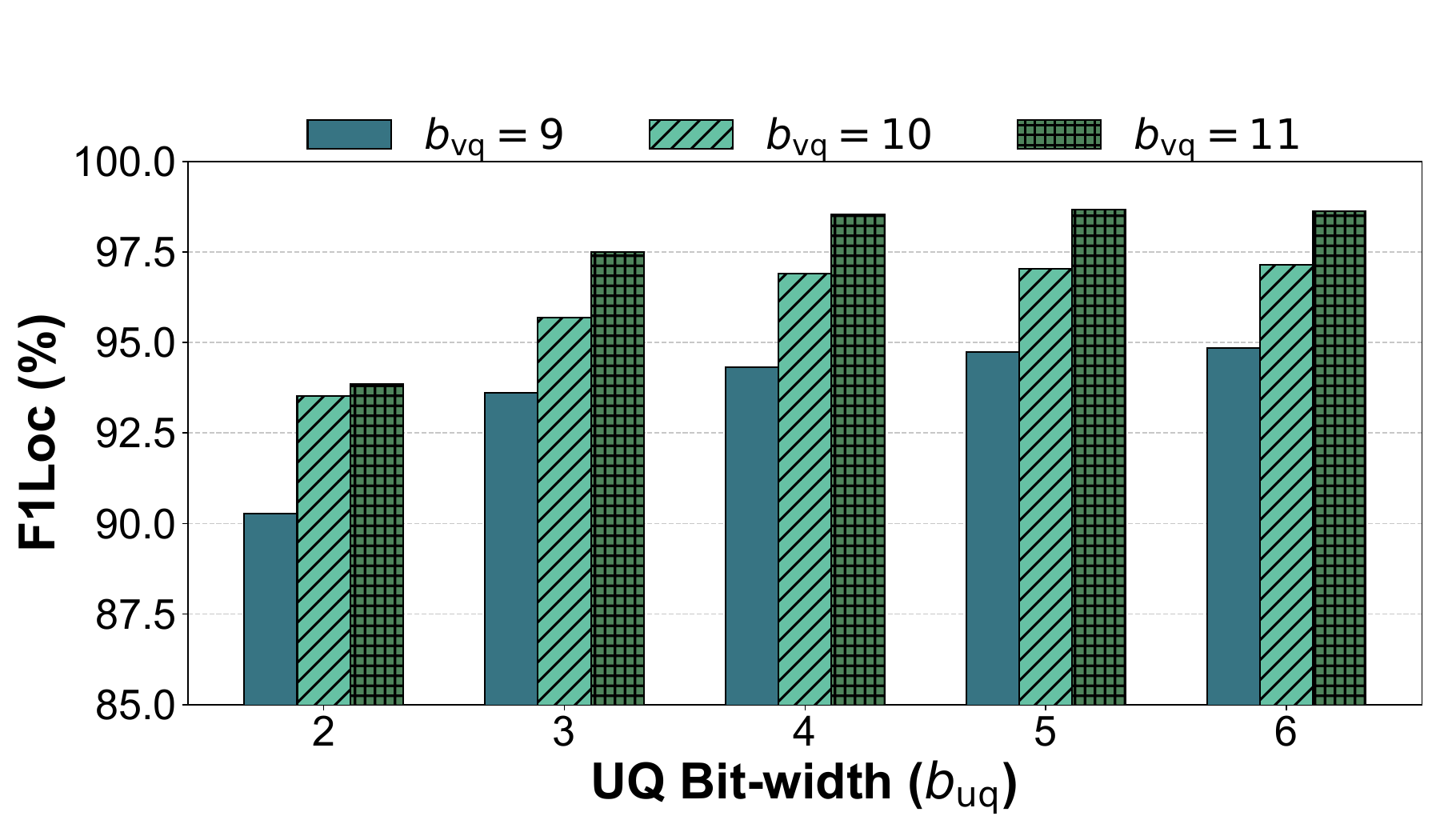}
		\label{fig:ex_4_d}
	}
	\caption{Impact of UQ codebook bit-width ($b_{\text{uq}}$) on Loc.}
	\label{fig_ex_4_2} 
\end{figure}

\noindent\textbf{Impact of $b_{\text{uq}}$ on UQ+VQ System.}
Next, we add a UQ before the VQ-only system, where the whole pipelined process of PCA$\to$UQ$\to$VQ gets implemented on each OPM to encode inputs for AE, and the VQ applied before each aggregation in GraphSAGE is still not used. To quantify the information loss incurred by the quantization on OPM, we select the bit-width of VQ codebook within ($b_{\text{vq}} \in [9, 11]$) while varying the bit-width of UQ ($b_{\text{uq}}$).

Figures~\ref{fig_ex_4_1} and~\ref{fig_ex_4_2} illustrate the UQ+VQ system's performance on anomaly classification and root-cause location, respectively. The performance on accuracies and F1-scores consistently improves with the increase of the UQ bit-width ($b_{\text{uq}}$). The improvements mainly occur when $b_{\text{uq}}$ increases from 2 to 4, and saturate beyond $b_{\text{uq}}=4$ bits. This suggests that a 4-bit quantization in the UQ can strike a balance between preserving essential feature information and effective data compression. Therefore, we select $b_{\text{uq}} \in [4, 6]$ below.

\noindent\textbf{Impact of $b_{\text{agg}}$ on Full Implementation.}
Finally, on top of the UQ+VQ system, we add a VQ before each aggregation in GraphSAGE to fully implement RIGEL. The experiments use $b_{\text{uq}} \in [4,6]$ and $b_{\text{vq}} \in [10,11]$, and vary the codebook bit-width of the VQ before each aggregation in GraphSAGE ($b_{\text{agg}}$) within $[6,7]$, \textit{i.e.}, the size of the corresponding codebook ($C_{\text{agg}}$) are 64 and 128, respectively. This choice is made based on the memory usage of the corresponding MAT. Specifically, as each key of the MAT contains an index pair from $C_{\text{agg}}$ (for the node and its neighbor, respectively), the MAT's size scales with $|C_{\text{agg}}|^2$. Hence, $b_{\text{agg}}=7$ becomes the maximum value that is feasible for our implementation.

Table~\ref{tab:hyperparam_results} details the performance of the full implementation of RIGEL. We notice that $b_{\text{agg}}$ acts as the main performance bottleneck, \textit{i.e.}, when the combination of $b_{\text{vq}}$ and $b_{\text{uq}}$ is fixed, increasing $b_{\text{agg}}$ from 6 to 7 leads to significant gains across all the metrics. For instance, this one-bit increase boosts F1Cls by up to $16.91\%$ (from $79.92\%$ to $96.83\%$ when $b_{\text{vq}} = 11$ and $b_{\text{uq}} = 5$), and improves F1Loc by up to $8.20\%$ (from $89.63\%$ to $97.83\%$ under the same setting).

Meanwhile, it is interesting to see that when the combination of $b_{\text{vq}}$ and $b_{\text{agg}}$ is fixed, increasing $b_{\text{uq}}$ does not always improve performance, especially when $b_{\text{agg}}=6$, \textit{e.g.}, with $b_{\text{vq}}=11$ and $b_{\text{agg}}=6$, increasing $b_{\text{uq}}$ from 4 to 5 causes F1Cls to drop from $94.46\%$ to $79.92\%$. This suggests that there is a complex interplay between UQ and VQ, which can make excessive input precision not beneficial when the VQ before each aggregation in GraphSAGE is the major information bottleneck. Finally, Table~\ref{tab:hyperparam_results} suggests an optimal configuration of the bit-widths as ($b_{\text{vq}}=11$, $b_{\text{uq}}=6$, and $b_{\text{agg}}=7$).

\begin{table*}[t!]
	\centering\footnotesize
	\caption{\centering Cross-Topology Generalization Capability of RIGEL.}
	\label{tab:cross_topology}
	\begin{tabular}{ll c c c c c}
		\toprule
		\textbf{Training Setup} & \textbf{Metrics (\%)} & \textbf{6-Node Topology} & \textbf{8-Node Topology} & \textbf{10-Node Topology} & \textbf{12-Node Topology} & \textbf{14-Node Topology} \\
		\midrule
		\multirow{2}{*}{\textbf{6-Node Topology}}  
		& AccCls / F1Cls & 99.44 / 98.04 & 99.08 / 95.36 & 99.13 / 94.99 & 99.08 / 93.34 & 98.94 / 90.73 \\
		& AccLoc / F1Loc & 99.51 / 98.32 & 99.24 / 96.46 & 99.19 / 95.21 & 99.19 / 94.22 & 99.03 / 91.78 \\
		\midrule
		\multirow{2}{*}{\textbf{14-Node Topology}} 
		& AccCls / F1Cls & 99.55 / 98.43 & 99.37 / 96.93 & 99.30 / 95.85 & 99.29 / 94.86 & 99.12 / 92.48 \\
		& AccLoc / F1Loc & 99.60 / 98.65 & 99.47 / 97.54 & 99.36 / 96.28 & 99.37 / 95.57 & 99.20 / 93.27 \\
		\bottomrule
	\end{tabular}
\end{table*}

\begin{table*}[t!]
\centering\footnotesize
\caption{\centering Detection of Unseen Anomalies via Multi-Root Reporting in Various Topologies.}
\label{tab:novel_faults}
\begin{tabular}{l c c c c c}
\toprule
\textbf{Metrics (\%)} & \textbf{6-Node Topology} & \textbf{8-Node Topology} & \textbf{10-Node Topology} & \textbf{12-Node Topology} & \textbf{14-Node Topology} \\ \midrule
Multi-Root Reporting (\textit{Class} 9 / \textit{Class} 10) & 93.0 / 89.6 & 99.3 / 98.9 & 100.0 / 100.0 & 100.0 / 100.0 & 99.8 / 99.7 \\
Multi-Root Reporting (Known Anomalies) & 0.08 & 0.00 & 0.20 & 0.60 & 0.48 \\ \bottomrule
\end{tabular}
\end{table*}

In all, the results above suggest that in practical deployments, the accuracy-efficiency tradeoff of RIGEL should be adjusted as follows. To improve diagnosis accuracy, an operator should first try to increase the VQ/UQ bit-widths, while adding more neighbors in each neighbor aggregation should always be the second choice. This is because feature exchange among more neighbors not only increases diagnosis latency but also necessitates reduction of the quantization bit-width allocated to each neighbor under the per-packet metadata budget of Tofino switch, degrading feature quality.

\subsection{Generalization of GraphSAGE Model}
\label{subsec:New Topologies and Novel Faults}

An important feature of ML-based anomaly diagnosis is the generalization of its model, \textit{i.e.}, whether the model can be applied to various networks and detect unseen anomalies, without requiring excessive retraining. In the following, we evaluate RIGEL's generalization over these two dimensions.

\noindent\textbf{Generalization over Topologies.} We first test the cross-topology generalization capability of the GraphSAGE model in RIGEL, by considering two topologies (\textit{i.e.}, the original 6-node topology in Figure~\ref{fig3a} and the 14-node NSFNET topology \cite{zhu2013nsfnet}) and conducting bi-directional transferability evaluations. First, we train the model in the 6-node topology and evaluate it in various topologies by adding two nodes each time until reaching the 14-node topology. Conversely, we train the model in the 14-node topology and evaluate it in smaller topologies down to the 6-node one. In all the scenarios, we directly deploy the GraphSAGE model with quantized weights without any fine-tuning.

Table~\ref{tab:cross_topology} shows the results. When the topologies used in training and testing are the same, all the performance metrics exceed 92\%, confirming that RIGEL can easily adapt to various topologies without performance degradation. The results from the cases in which the training and testing topologies are different verify that exceptional cross-topology generalization can be achieved. For example, when we apply the model trained in the 6-node topology to the 14-node one, AccCls and F1Loc reach 98.94\% and 90.73\%, respectively. It is interesting to see that applying the model trained in the 14-node topology to a smaller topology can even achieve metrics higher than those obtained in the 14-node one. This is because training the model in a larger topology makes it observe a more diverse set of fault propagation paths and deeper cascading effects, yielding superior backward compatibility.

\noindent\textbf{Generalization for Unseen Anomalies.} 
While RIGEL can achieve high accuracy on detecting known anomalies, we evaluate how well it performs when there are unseen anomalies (a practical challenge in real-world optical networks). Specifically, we expose the deployed GraphSAGE model to two unseen anomalies: \textit{Class} 9 (a BV-WSS center frequency drifting of 37.5 GHz) and \textit{Class} 10 (a fault that combines spectrum narrowing and narrow-band noise insertion).

\begin{table*}[t!]
	\centering\footnotesize
	\caption{\centering Comparisons of Model Performance and Bandwidth Overhead.}
	\label{tab:comm_overhead}
	\begin{tabular}{l S[table-format=2] S[table-format=2] S[table-format=2] S[table-format=2] c c}
		\toprule
		Model & {AccCls (\%)} & {AccLoc (\%)} & {F1Cls (\%)} & {F1Loc (\%)} & {Interactions} & {\makecell{Bandwidth \\ Overhead (bits)}} \\
		\midrule
		C-GNN         & 99.99 & 99.99 & 99.99 & 99.99 & 66,950 & 266,728,800 \\
		C-GNN-UQ & 99.65 & 99.65 & 98.86 & 98.82 & 66,950 & 57,844,800
 \\
		RIGEL               & 99.42 & 99.55 & 98.08 & 98.34 & 6,490  & 103,840
 \\
		\bottomrule
	\end{tabular}
\end{table*}

Table~\ref{tab:novel_faults} summarizes the results, where each model is originally trained on a topology with $[6,14]$ nodes and anomalies in \textit{Classes} 0-8. We can see that when the unseen anomalies occur, the ratio of multi-root reporting is always 98.9\% or higher, except for the cases with the 6-node topology, where the ratio can drop to 89.6\%. The noticeable ratio drop in the 6-node topology is caused by its small topology size, \textit{i.e.}, the percentage of short lightpaths in it is much larger than that in other larger topologies, which hides certain cases of multi-root reporting. On the other hand, when there is no unseen anomaly, the ratio of multi-root reporting is always 0.6\% or smaller. The significant difference in the ratio of multi-root reporting verifies that it can be used as a good indicator of unseen anomalies to the control plane, \textit{i.e.}, RIGEL can reliably detect the occurrences of unseen anomalies and trigger timely retraining, effectively closing the autonomous diagnostic loop without much human intervention.

\subsection{Comparisons with Centralized Diagnosis}
\label{subsec:comparisons}

Finally, we compare RIGEL with the conventional optical anomaly diagnosis schemes that use centralized data analytics in the control plane (based on centralized GraphSAGE-based GNNs). The optimal configuration ($b_{\text{vq}}=11$, $b_{\text{uq}}=6$, and $b_{\text{agg}}=7$) is used for RIGEL. The schemes using a centralized GNN (C-GNN) and a centralized GNN with UQ (C-GNN-UQ) ($b_{\text{uq}}=6$) are the baselines, and they both use the full-precision model for anomaly diagnosis since an SDN controller usually does not have computation restrictions. C-GNN makes each optical node report the 20-dimensional real data compressed by the PCA in each OPM, which contributes $20 \times 32=640$ bits to each payload of a telemetry packet. C-GNN-UQ lets each node report the 20-dimensional integer vectors at the UQ's output, occupying $20 \times 6 =120$ bits in each payload. The experiments use the 6-node topology and focus on two metrics: the interactions with the control plane and the total bandwidth overhead. 

For the centralized baselines, each reporting cycle involves every optical node sending a telemetry packet to the control plane. In contrast, the reporting in RIGEL is event-driven, and thus only one control message, whose payload only occupies $16$ bits on encoding the diagnosis result (\textit{i.e.}, the impacted lightpath, the class of its anomaly and location), is sent to the control plane, after detecting an anomaly. Table \ref{tab:comm_overhead} lists the results when we run experiments over all the $\sim$$400,000$ testing samples. The interactions made by the two baselines are the same, while RIGEL reduces the interactions by $10.3$$\times$. The bandwidth overhead reductions by RIGEL are more significant, which are $\sim$$2,568$$\times$ and $\sim$$557$$\times$ relative to C-GNN and C-GNN-UQ, respectively. Meanwhile, we notice that the accuracies and F1-scores of RIGEL are slightly lower than those of the benchmarks (the reductions are all within $2\%$). This is because it adopts the quantized AE+GraphSAGE. The minor performance degradation is acceptable, since the accuracies and F1-scores of RIGEL are all above $98\%$ for effective optical anomaly diagnosis.

\section{Discussions and Related Work}

\noindent\textbf{Methodological Advancement over Prior Art.} Previous NN-on-switch systems, notably Quark~\cite{zhang2025quark} and Brain-on-Switch~\cite{yan2024brain}, made significant progress by adapting NNs to switch hardware. However, they were primarily based on UQ, leading to costly per-dimension iterative processing in PDP pipeline. In contrast, our software-hardware co-design of UQ+VQ maps each sample of high-dimensional telemetry data to a single VQ index for MAT lookup. This bypasses the bottleneck due to per-dimension processing, unlocking a collaborative inference capability that is absent in both DT-based approaches~\cite{butun2025dune, xiong2019switches} and software-based GNNs~\cite{silva2023confidentiality, chen2022cooperative}.

\noindent\textbf{A Unified Solution to Hardware Challenges.} The core innovation of our work lies in how we use VQ index in RIGEL, which simultaneously resolves the two major hardware challenges of realizing on-switch GNNs, \textit{i.e.}, in-switch processing and inter-switch communication. As for the first one, the VQ index realizes an efficient key for MAT lookups, solving the memory explosion problem~\cite{qian2019flexgate}. As for the second one, the VQ index enables low-overhead neighbor aggregation, effectively reducing inter-switch communication overheads. To the best of our knowledge, this unified solution is the first to enable a distributed in-network diagnosis system.

\section{Conclusion}

In this paper, we presented RIGEL, a distributed GNN-based framework for fully in-network optical anomaly diagnosis. Our key contribution is a VQ-based quantization mechanism to transform a complex GNN for optical anomaly diagnosis into a series of MATs that can be easily offloaded on Tofino switch. We prototyped RIGEL and evaluated it thoroughly in a realistic packet-over-optical network testbed for real-time anomaly classification and location in the optical layer. The results demonstrated that RIGEL realizes accurate anomaly diagnosis, and achieves more than three orders of magnitude reduction in communication overheads between data and control planes, over the conventional schemes with centralized data analytics.

\bibliographystyle{plain}
\bibliography{ref_response}

\clearpage
\appendix
\section{Appendix}

\subsection{Codebook-based Discretization}
\label{sec:VQ}
\begin{figure}[t] 
	\centering 
	\includegraphics[width=0.82\columnwidth]{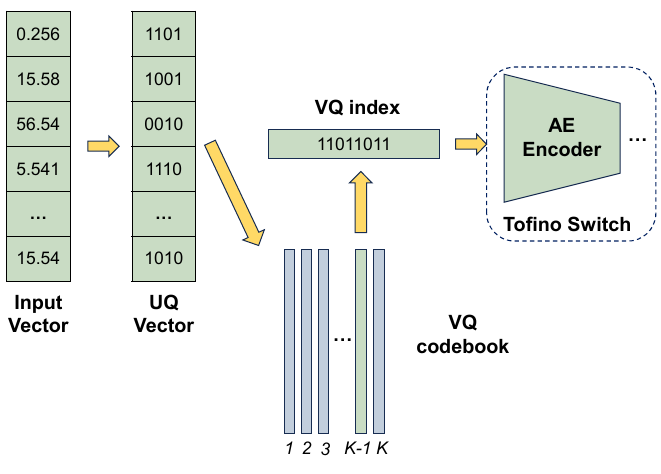}
	\caption{UQ-VQ feature discretization process.}
	\label{fig5} 
\end{figure}

Although VQ follows UQ in the pipeline in Figure~\ref{fig5}, the design and training of the UQ are actually based on the trained VQ codebook. Hence, we first explain the design of the VQ. 

We discretize features with a learned codebook at the following places: before the encoder in AE and before each aggregation for neighborhood fusion in GraphSAGE. Then, the discretized processing in AE and GraphSAGE is unified around VQ-codebook indices, which are MAT-friendly, saving the effort in modifying the AE and GraphSAGE to adapt to non-unified inputs. Specifically, as we encode each input (a VQ codebook index) with a fixed $b_{\text{vq}}$-bit format and do not modify its encoder architecture when offloading the AE, every intermediate layer in the encoder becomes a deterministic mapping from finite inputs to finite outputs, which can be easily realized with MAT lookups. This explains the rationale for placing a VQ before the encoder in AE, \textit{i.e.}, each sample of spectral data is first mapped to a learned codeword (a VQ codebook index) and then passed through the encoder, stabilizing the input distribution seen by each layer in the encoder and eliminating per-dimension loops. On the other hand, for the GraphSAGE, we also apply a VQ before each aggregation, where the self-feature and neighbor-feature are separately transformed into VQ codebook indices before being combined, to control the numeric dispersion induced by neighborhood propagation. This enables MAT-driven fusion over codewords, which reduces arithmetic operations and thus aligns with the capability of PDP.

Based on the design above, the codebooks for AE are first trained by aligning to the full-precision encoder. Let $E$ be the full-precision encoder, which is fixed during the training. For an input $x$ (high-dimensional spectral data), we define $z^\star = E(x)$ as its encoding result from $E$, and denote the VQ codebook as $C_{in}$. The input $x$ is mapped to its nearest codeword $c_{\text{in}} \in C_{\text{in}}$ by first finding the corresponding index:
\begin{equation}
\small
k_{\text{in}}(x)
  = \arg\min_{k}
    \bigl| x - C_{\text{in}}[k] \bigr|^{2}.
    \label{eq:kin}
\end{equation}
Thus, the codeword is $c_{\text{in}} = C_{\text{in}}[k_{\text{in}}(x)]$, and by putting $c_{\text{in}}$ in $E$, we get $z = E(c_{\text{in}})$. The loss between $z^\star$ and $z$ consists of a feature-alignment term with two lightweight regularizers:

\begin{equation}
\small
\label{eq:ae-vq-train}
\begin{split}
    \mathcal{L}_{\text{AE-VQ}} ={}& \underbrace{\lambda_{z} \cdot \!\left[1-\cos(z, z^\star)\right] + \mu_{z}\,\frac{\lVert z - z^\star\rVert_2^2}{\lVert z^\star\rVert_2^2+\varepsilon}}_{\text{feature alignment (cosine + NMSE)}} \\
    & + \underbrace{\lambda_{u}\cdot\mathcal{R}_{\text{usage}}(C_{\text{in}})}_{\text{codeword usage}} \\
    & + \underbrace{\lambda_{s}\cdot\lVert x_{\text{soft}}-x_{\text{soft}}^{\,(\text{noise})}\rVert_2^2}_{\text{stability}},
\end{split}
\end{equation}
where $\lambda{z}$, $\mu_{z}$, $\lambda_{u}$, and $\lambda_{s}$ are the non-negative hyper-parameters that balance the contribution of each term and are tuned on a validation set, $\varepsilon$ is a small constant (\textit{e.g.}, $10^{-8}$) added to prevent division by zero in the NMSE calculation, and the regularizer $\mathcal{R}_{\text{usage}}$ discourages codeword collapse by promoting uniform codebook utilization:
\begin{equation}
\small
\begin{aligned}
    &\mathcal{R}_{\text{usage}}(C_{\text{in}}) = \left[\log \left(K_{\text{in}}\right) - H(p_{\text{in}})\right]^2, \\
    &H(p_{\text{in}}) = -\sum_{k=1}^{K_{\text{in}}} p_{\text{in}}(k)\cdot\log \left[p_{\text{in}}(k)\right], \label{eq:usage}
\end{aligned}
\end{equation}
where $p_{\text{in}}(k)$ is the empirical usage frequency of codeword $k$ within a batch, and $K_{\text{in}}$ is the size of the codebook $C_{\text{in}}$.

The stability term in Eq. \eqref{eq:ae-vq-train} enforces a robust mapping by penalizing sensitivity to small input perturbations. It compares two softly-reconstructed inputs, $x_{\text{soft}}$ and $x_{\text{soft}}^{(\text{noise})}$. The soft reconstruction is a differentiable proxy for the quantization process, \textit{i.e.}, $x_{\text{soft}}$ is calculated as a weighted average of all the codewords, where the weights are derived by applying a softmax function to the distances between the input $x$ and each codeword in $C_{\text{in}}$. $x_{\text{soft}}^{(\text{noise})}$ is calculated similarly, but with a slightly perturbed input $(x+\text{noise})$.

To obtain the codebooks for GraphSAGE, we still leverage a trained full-precision model of GraphSAGE. Let $\ell^\star$ be the logits from the full-precision model, and $\ell$ be the logits from the discretized GraphSAGE (\textit{i.e.}, that whose internal features are quantized). Here, logits are the scores from a GNN's final layer. Then, the loss used in our implementation becomes
\begin{equation}
\label{eq:gnn-vq-train}
\small
\begin{split}
\mathcal{L}_{\text{GNN-VQ}} 
  &= \underbrace{\mathrm{CE}(\ell_{\text{cls}}, y_{\text{cls}})
     + \gamma\cdot\mathrm{BCE}(\ell_{\text{loc}}, y_{\text{loc}})}_{\text{task supervision}} \\
  &\quad + \eta\,\underbrace{\Bigl[\alpha\cdot\operatorname{NMSE}(\ell,\ell^\star)
     + (1-\alpha)\bigl(1-\cos(\ell,\ell^\star)\bigr)\Bigr]}_{\text{logits alignment (cosine + NMSE)}} \\
  &\quad + \underbrace{\lambda_{u}\cdot\mathcal{R}_{\text{usage}}}_{\text{codebook usage regularization}},
\end{split}
\end{equation}
where $\alpha$ is the mixing coefficient inside the cosine-NMSE alignment, $\gamma$ balances the classification and location heads, and $\eta$ controls the relative weight of the logits alignment term (set to $0.1$ in our implementation, but treated as a tunable hyperparameter). The task supervision term employs cross-entropy (CE) for the multi-class fault classification head ($\ell_{\text{cls}}$) and binary cross-entropy (BCE) for the location head ($\ell_{\text{loc}}$). The logits alignment term distills knowledge from the full-precision GraphSAGE. Finally, the codebook usage regularizer $\mathcal{R}_{\text{usage}}$ prevents codebook collapse. 

\textit{Algorithm}~\ref{alg:vq-multiphase} outlines our multi-step training pipeline for VQ to get the codebooks for AE and GraphSAGE. Its inputs are the sets of spectral data ($X$) and corresponding labels ($Y$), sizes of VQ codebooks for AE ($K_{\text{in}}$) and GraphSAGE ($K_{\text{agg}}$), and related hyperparameters, and it outputs the full-precision models for AE ($E$) and GraphSAGE ($G$), discretized GraphSAGE ($G_{\text{vq}}$), and the codebooks for AE ($C_{\text{in}}$) and GraphSAGE ($C_{\text{agg}}$). \textbf{Step 0} (\textit{Lines} 1-3) trains the full-precision models of AE and GraphSAGE. \textbf{Step 1} (\textit{Lines} 4-7) learns the codebook of AE ($C_{\text{in}}$). Specifically, in each iteration, we refine the codebook by minimizing $\mathcal{L}_{\text{AE-VQ}}$ to align latent vectors $z$ and $z^\star$, and also update $C_{\text{in}}$ via exponential moving average (EMA). \textbf{Step 2} (\textit{Lines} 8-11) learns the discretized GraphSAGE $G_{\text{vq}}$ and the pre-aggregation codebook $C_{\text{agg}}$. Specifically, after the initialization with K-Means (\textit{Line} 8), $C_{\text{agg}}$ is updated via EMA while minimizing $\mathcal{L}_{\text{GNN-VQ}}$, which enforces task supervision and logits-level distillation ($\ell$ \textit{versus} $\ell^\star$). \textit{Line} 12 returns $E$, $G$, $G_{\text{vq}}$, $C_{\text{in}}$, and $C_{\text{agg}}$.

\begin{algorithm}[!htb]
\caption{Multi-Step Training with VQ}
\label{alg:vq-multiphase}

\KwIn{$X$, $Y$, $K_{\text{in}}$, $K_{\text{agg}}$.}
\BlankLine
\tcc{Step0: Train full-precision models of AE and GraphSAGE}
\For{each epoch}{
    train an AE on $X$ via MSE reconstruction loss to get full-precision encoder $E$\;
    use $\{z\}$ as inputs to train a GraphSAGE $G$, with loss as the task supervision term in Eq.~\eqref{eq:gnn-vq-train}\;
}
\BlankLine
\tcc{Step1: VQ distillation of AE}
initialize $C_{\text{in}}$ by taking a random subset from the first batch of $N$ samples from $X$ to form $K_{\text{in}}$ initial indices (or repeating copies if $N<K_{\text{in}}$)\;\vspace{2pt}

\For{each epoch}{
    update $C_{\text{in}}$ by minimizing $\mathcal{L}_{\text{AE-VQ}}$ in Eq.~\eqref{eq:ae-vq-train}\;
    update $C_{\text{in}}$ with EMA\;
}

\BlankLine
\tcc{Step2: VQ distillation of GraphSAGE}
initialize $C_{\text{agg}}$ via K-Means on internal layer before each aggregation in $G$\;

\For{each epoch}{
    train $G_{\text{vq}}$ and $C_{\text{agg}}$ with $X$ and $Y$ to minimize composite loss $\mathcal{L}_{\text{GNN-VQ}}$ in Eq.~\eqref{eq:gnn-vq-train}\;
    update $C_{\text{agg}}$ with EMA\;
}

\Return{$E, G, G_{\text{vq}},C_{\text{in}}, C_{\text{agg}}$}
\end{algorithm}

\subsection{Input Quantization with UQ}
\label{sec:UQ}
We leverage per-dimension UQ to discretize the spectral data in $X$, where each sample is $x \in \mathbb{R}^D$ (a $D$-dimensional vector). For each $x$, we specify a bit-width $b_{\text{uq}}$, yielding a codeword length of $K=2^{b_{\text{uq}}}$. This scheme is parameterized by a learnable step size $\Delta_d$ and an offset (zero-point) $\sigma_d$ for each dimension $d$. The mapping of UQ involves two steps, where the parameters $\Delta_d$ and $\sigma_d$ are initialized from a small batch in $X$ by mapping each dimension’s empirical range to the integer interval $[0, K-1]$. Then, the two steps works as:
\begin{equation}
\small
\left\{
\begin{aligned}
    q_d &= \operatorname{clip}\!\left[
        \operatorname{round}\!\Bigl(\tfrac{x_d - \sigma_{d}}{\Delta_d}\Bigr),
        0, K-1 \right] \ \ \text{(Step 1)},\\
    \hat{x}_d &= \Delta_d \cdot q_d + \sigma_{d} \ \ \text{(Step 2)}.
\end{aligned}
\right.
\end{equation}
First, the floating-point value of dimension $d$ in $x$ ($x_d$) is mapped to an integer $q_d \in [0, K-1]$. This involves rounding the scaled and shifted value of $x_d$, followed by a clipping operation to ensure that $q_d$ stays within the valid range. Second, the obtained integer $q_d$ is used to compute the de-quantized value $\hat{x}_d$. After getting $\hat{x}_d$ for each $x_d$, we obtain the reconstructed vector $\hat{x}$, which will be passed to the AE's encoder. 

The training of the UQ is essentially to update its parameters $\{\Delta_d, \sigma_d\}$ under the guidance of a composite loss function, $\mathcal{L}_{\text{UQ}}$, which aligns the UQ reconstruction with both the input-space VQ structure and the latent-space geometry.
\begin{equation}
\label{eq:uq-train}
\small
\begin{aligned}
\mathcal{L}_{\text{UQ}}
  &= \underbrace{\bigl\lVert \hat{x} - c_{\text{in}}\bigr] \bigr\rVert_2^2}_{\text{UQ-VQ Alignment}} \\
  &\quad + \lambda_{\text{feat}} \Bigl[ \alpha \cdot \operatorname{NMSE}\!\bigl(E(\hat{x}),\,E(x)\bigr) \\
  &\quad + (1-\alpha) \cdot \Bigl(1 - \cos\!\bigl(E(\hat{x}),\,E(x)\bigr)\Bigr) \Bigr],
\end{aligned}
\end{equation}
where $\lambda_{\text{feat}}>0$ is the weight of the latent alignment term. The training procedure optimizes $\{\Delta_d, \sigma_d\}$ to create a stable mapping from $x_d$ (a real number) to $q_d$ (an integer), ensuring the quantized results to align with the subsequent processing.
\end{document}